\documentclass[fleqn,usenatbib]{mnras}

\usepackage{newtxtext,newtxmath}
\usepackage[T1]{fontenc}

\DeclareRobustCommand{\VAN}[3]{#2}
\let\VANthebibliography\thebibliography
\def\thebibliography{\DeclareRobustCommand{\VAN}[3]{##3}\VANthebibliography}

\usepackage{graphicx}	% Including figure files
\usepackage{amsmath}	% Advanced maths commands
\usepackage{float}
\usepackage{textcomp}
\usepackage{placeins}
\usepackage{multirow}
\usepackage{latexsym}
\usepackage{threeparttable}
\usepackage{longtable}
\usepackage{tabularx}
\usepackage{lscape}
\usepackage{ulem} % Added by JK for strikeout

\hypersetup{
     colorlinks   = true,
     citecolor     = blue
} 

\providecommand{\bjdtdb}{\ensuremath{\rm {BJD_{TDB}}}}

\providecommand{\teff}{\ensuremath{T_{\rm eff}}}

\providecommand{\msun}{\ensuremath{\,M_{\odot}}}
\providecommand{\rsun}{\ensuremath{\,R_{\odot}}}
\providecommand{\lsun}{\ensuremath{\,L_{\odot}}}
\providecommand{\mj}{\ensuremath{\,M_{\rm J}}}
\providecommand{\rj}{\ensuremath{\,R_{\rm J}}}
\providecommand{\me}{\ensuremath{\,M_{\rm \oplus}}}
\providecommand{\re}{\ensuremath{\,R_{\rm \oplus}}}
\providecommand{\fave}{\langle F \rangle}
\providecommand{\fluxcgs}{10$^9$ erg s$^{-1}$ cm$^{-2}$}
\providecommand{\densitycgs}{g\,cm$^{-3}$}
\providecommand{\logg}{cm\,s$^{-2}$}

\newcommand\mm{\mbox{$\mu$m}}% 
\newcommand{\mos}{\,m\,s$^{-1}$}
\newcommand{\kms}{\,km\,s$^{-1}$}

\newcommand\msini{\ifmmode{{\mathrm M} \sin i}\else${{\mathrm M} \sin i}$\fi}

\newcommand{\Mearth}{\mbox{$M_{\oplus}$}}
\newcommand{\Rearth}{\mbox{$R_{\oplus}$}}

\newcommand{\muHz}{\mbox{$\mu$Hz}}

\newcommand{\numax}{\ensuremath{\nu_{\textrm{max}}}}
\newcommand{\dnu}{\ensuremath{\Delta\nu}}
\newcommand{\kep}{\mbox{\textit{Kepler}}}
\newcommand{\target}{TOI-257}

\newcommand{\rhostar}{\mbox{$\rho_{\star}$}}
\newcommand{\radstar}{\mbox{$1.888\pm0.033$}}
\newcommand{\massstar}{\mbox{$1.390\pm0.046$}}
\newcommand{\agestar}{\mbox{$3.46\pm0.43$}}
\newcommand{\loggstar}{\mbox{$4.030\pm0.011$}}
\newcommand{\denstar}{\mbox{$0.293\pm0.011$}}
\newcommand{\mstar}{\mbox{$M_{\star}$}}

\newcommand{\rstar}{\mbox{$R_{\star}$}}

\newcommand{\teffstar}{\mbox{$6075 \pm 90$}}
\newcommand{\fehstar}{\mbox{$0.19 \pm 0.10$}}

\title{TOI-257b (HD 19916b): A Warm sub-Saturn Orbiting an Evolved F-type Star}

\author[B. C. Addison et al.]{
\parbox{\textwidth}{
\large Brett C. Addison,$^{1}$\thanks{E-mail: \href{Brett.Addison@usq.edu.au}{Brett.Addison@usq.edu.au}}
Duncan J. Wright,$^{1}$
Belinda A. Nicholson,$^{2,1}$
Bryson Cale,$^{3}$
Teo Mocnik,$^{4}$
Daniel Huber,$^{5}$
Peter Plavchan,$^{3}$ %\newauthor \large
Robert A. Wittenmyer,$^{1}$
Andrew Vanderburg,$^{6}$
William J. Chaplin,$^{7,8}$
Ashley Chontos,$^{5,9}$
Jake T. Clark, $^{1}$
Jason D. Eastman,$^{10}$ %\newauthor \large
Carl Ziegler,$^{11}$
Rafael Brahm,$^{12,13}$
Bradley D. Carter,$^{1}$
Mathieu Clerte,$^{1}$
N\'estor Espinoza,$^{14}$
Jonathan Horner,$^{1}$ %\newauthor \large
John Bentley,$^{32}$
Andr\'es Jord\'an,$^{16,13}$
Stephen R. Kane,$^{4}$
John F. Kielkopf,$^{17}$
Emilie Laychock,$^{18}$
Matthew W. Mengel,$^{1}$
Jack Okumura,$^{1}$ %\newauthor \large
Keivan G.\ Stassun,$^{19,20}$
Timothy R. Bedding,$^{21,8}$
Brendan P. Bowler,$^{22}$
Andrius Burnelis,$^{30}$
Sergi Blanco-Cuaresma,$^{10}$
Michaela Collins,$^{18}$ %\newauthor \large
Ian Crossfield,$^{23,24}$
Allen B. Davis,$^{25}$
Dag Evensberget,$^{1}$
Alexis Heitzmann,$^{1}$
Steve B. Howell,$^{27}$
Nicholas Law,$^{28}$
Andrew W. Mann,$^{28}$ %\newauthor \large
Stephen C. Marsden,$^{1}$
Rachel A.~Matson,$^{26}$
James O'Connor,$^{1}$
Avi Shporer,$^{31}$
Catherine Stevens,$^{30}$
C.G. Tinney,$^{32}$
Christopher Tylor,$^{1}$
Songhu Wang,$^{15}$ %\newauthor \large
Hui Zhang,$^{33}$
Thomas Henning,$^{34}$
Diana Kossakowski,$^{34}$
George Ricker,$^{31,62}$
Paula Sarkis,$^{34}$
Martin Schlecker,$^{34}$
Pascal Torres,$^{29}$
Roland Vanderspek,$^{31}$
David W. Latham,$^{10}$
Sara Seager,$^{31,35}$ %\newauthor \large
Joshua N.\ Winn,$^{36}$
Jon M. Jenkins,$^{37}$
Ismael Mireles,$^{31}$
Pam Rowden,$^{38}$
Joshua Pepper,$^{39}$
Tansu Daylan,$^{31}$
Joshua E. Schlieder,$^{40}$
Karen A.\ Collins,$^{10}$
Kevin I.\ Collins,$^{3}$
Thiam-Guan Tan,$^{41}$
Warrick H. Ball,$^{7,8}$
Sarbani Basu,$^{25}$ %\newauthor \large
Derek L. Buzasi,$^{43}$
Tiago L. Campante,$^{44,45}$
Enrico Corsaro,$^{46}$
Luc\'ia Gonz\'alez-Cuesta,$^{47,48}$ %\newauthor \large
Guy R. Davies,$^{7,8}$
Leandro~de~Almeida,$^{49}$
Jose-Dias~do~Nascimento,~Jr.,$^{49,10}$
Rafael ~A.~Garc\'\i a,$^{50,51}$
Zhao Guo,$^{52}$ %\newauthor \large
Rasmus Handberg,$^{8}$
Saskia Hekker,$^{60,61,8}$
Daniel R.\ Hey,$^{21,8}$
Thomas Kallinger,$^{54}$
Steven D. Kawaler,$^{55}$
Cenk Kayhan,$^{56}$ %\newauthor \large
James S.\ Kuszlewicz,$^{53,8}$
Mikkel N.\ Lund,$^{8}$
Alexander Lyttle,$^{7,8}$
Savita Mathur,$^{47,48}$
Andrea Miglio,$^{7,8}$
Benoit Mosser,$^{57}$ %\newauthor \large
Martin B. Nielsen,$^{7,8,42}$
Aldo M. Serenelli,$^{58,59}$
Victor Silva Aguirre,$^{8}$
\& Nathalie Theme\ss l$^{53,8}$
}
\vspace{0.4cm}
\\
\parbox{\textwidth}{
The authors' affiliations are shown in Appendix \ref{sec:affiliations}.}}

\date{Accepted XXX. Received YYY; in original form ZZZ}

\pubyear{2020}

\begin{document}
\label{firstpage}
\pagerange{\pageref{firstpage}--\pageref{lastpage}}
\maketitle

\begin{abstract}
We report the discovery of a warm sub-Saturn, TOI-257b (HD 19916b), based on data from NASA's Transiting Exoplanet Survey Satellite ({\textit {TESS}}). The transit signal was detected by {\textit {TESS}} and confirmed to be of planetary origin based on radial velocity observations. An analysis of the {\textit {TESS}} photometry, the {\sc {\textsc{Minerva}}}-Australis, FEROS, and HARPS radial velocities, and the asteroseismic data of the stellar oscillations reveals that TOI-257b has a mass of $M_P=0.138\pm0.023$\,$\rm{M_J}$ ($43.9\pm7.3$\,\me), a radius of $R_P=0.639\pm0.013$\,$\rm{R_J}$ ($7.16\pm0.15$\,\re), bulk density of $0.65^{+0.12}_{-0.11}$ (cgs), and period $18.38818^{+0.00085}_{-0.00084}$\,$\rm{days}$. TOI-257b orbits a bright ($\mathrm{V}=7.612$\,mag) somewhat evolved late F-type star with $M_*=1.390\pm0.046$\,$\rm{M_{sun}}$, $R_*=1.888\pm0.033$\,$\rm{R_{sun}}$, $T_{\rm eff}=6075\pm90$\,$\rm{K}$, and $v\sin{i}=11.3\pm0.5$\kms. Additionally, we find hints for a second non-transiting sub-Saturn mass planet on a $\sim71$\,day orbit using the radial velocity data. This system joins the ranks of a small number of exoplanet host stars ($\sim100$) that have been characterized with asteroseismology. Warm sub-Saturns are rare in the known sample of exoplanets, and thus the discovery of TOI-257b is important in the context of future work studying the formation and migration history of similar planetary systems.
\end{abstract}


%%%%%%%%%%%%%%%%%%%%%%%%%%%%%%%%%%%%%%%%%%%%%%%%%%

%%%%%%%%%%%%%%%%% BODY OF PAPER %%%%%%%%%%%%%%%%%%

\section{Introduction}
\label{intro}
When \citet{mayor1995} announced the discovery of the first hot Jupiter, 51 Pegasi b, astronomers were baffled by the existence of a Jovian planet orbiting its host star with such a short orbital period (about 4.2\,days). That discovery revolutionized our understanding of the planet formation process, revealing the situation to be more complex than had been expected based on studies of the Solar System \citep[e.g.,][]{lissauer1993}. Radial velocity and transit surveys over the past two decades have uncovered numerous warm and hot giant exoplanets with orbital periods shorter than 100 days \citep[see, e.g., ][]{butler1997,bayliss2013,brahm2016,2018MNRAS.478.4866V,dawson2019,kipping2019}, and occurrence studies based on those discoveries suggest that such planets can be found orbiting $\sim1\%$ of all Sun-like stars \citep[e.g.,][]{2010Sci...330..653H,2012ApJS..201...15H,2012A&A...545A..76S,HJabund,2016A&A...587A..64S,zhou2019} (in comparison to an occurrence rate of at least $7\%$ for more distant planets; see, e.g., \citealt{2016AJ....152..206F,2020MNRAS.492..377W}).

In addition to the Solar System lacking a hot Jupiter, it also lacks other broad classes of planets such as super-Earths and mini-Neptunes ($\sim1.5-3$\,$R_{\oplus}$) as well as planets larger than Neptune and smaller than Saturn, known as sub-Saturns (which we have defined as planets with a radius between $\sim5-8$\,$R_{\oplus}$). Sub-Saturns are a key class of planets to study for understanding the formation, migration, and compositions of giant planets in general. Their large size requires a significant H/He envelope that comprises a majority of their planetary volume, yet their masses are sufficiently small that their cores are not degenerate (unlike for planets near the mass of Jupiter). This means that modeling the interiors of sub-Saturns can be simplified as a planet consisting of a high-density core surrounded by extended H/He envelope and where measurements of mass and radius enable a single family of solutions for the planet's core and envelope mass fraction \citep[e.g.][]{weiss14, petigura16, pepper17, petigura17}.

It is commonly thought that close-in giant planets, such as hot/warm Jupiters and sub-Saturns, do not form {\it in situ}, but instead originate beyond the protostellar ice line (typically located at several astronomical units from the host star) where there is sufficient solid material available to build up $\sim5-20M_{\oplus}$ cores \citep{pollack1996,Weidenschilling2005,rafikov2006}. In the case of Jovian planets, once their cores reach this critical mass regime, they begin to rapidly accrete gas from the protoplanetary disk to form their gaseous envelopes. This process continues until the disk is dispersed \citep{rafikov2006,tanigawa2007}, resulting in Jupiter-sized planets with masses of $\sim100-10,000$\,$M_{\oplus}$. For sub-Saturns, however, the runaway accretion of gas appears to either not have occurred at all or did occur but in a gas-depleted disk \citep{2018MNRAS.476.2199L}. As a result, sub-Saturns have masses that range from $\sim10$ to $100$\,$M_{\oplus}$. The mass of a sub-Saturn is strongly correlated with the metallicity of its host star, but is uncorrelated with the resulting radial size \citep{petigura17}. 

The sample of measurements for longer period ($P\geq10$\,d) `warm' giants and sub-Saturns thus far is small. The detection of more of these systems is then important to better constrain the formation and migration mechanisms of close-in planets.

One such source of warm giant planetary systems is NASA's \textit{Transiting Exoplanet Survey Satellite} \citep[{\textit {TESS}},][]{ricker2015}, launched on 18th April, 2018. As of 6th November, 2019, the \textit{TESS} mission has delivered a total of 1361 planetary candidates -- objects that require further observations from ground-based facilities to confirm the existence of the candidate exoplanets\footnote{Data from the NASA Exoplanet Archive, 6th November 2019}. To date, such follow-up observations have resulted in a total of 34 confirmed planetary discoveries \citep[e.g.][]{HD1397,2019ApJ...881L..19V,Quinn19,2019AJ....157...51W} -- and it is likely that many more planets will be confirmed in the months to come.

During its initial two-year primary mission, {\it TESS}\, is expected to discover several dozen warm Jupiters, Saturns, and sub-Saturns orbiting bright ($V<10$\,mag) stars \citep{sullivan2015,2018ApJS..239....2B,huang2018}. Those planets will be suitable targets for follow-up observations to measure their masses, through radial velocity measurements, to probe their atmospheric compositions, through transmission and emission spectroscopy, and to determine their spin-orbit angles through measurements of the Rossiter-McLaughlin effect.

In this work, we report the discovery of one such planet, TOI-257b (HD 19916b), based on photometric data obtained by \textit{TESS}, and follow-up observations using the {\textsc{Minerva}}-Australis facility at the University of Southern Queensland's Mt. Kent Observatory \citep{2018arXiv180609282W,addison2019}, the FEROS instrument \citep[$R=48,000$,][]{1999Msngr..95....8K} on the MPG 2.2\,m telescope at La Silla Observatory, and the HARPS spectrograph \citep[$R=120,000$,][]{2003Msngr.114...20M} on the ESO 3.6\,m telescope at La Silla Observatory. The details of the spectrographs and spectroscopic observations are provided in Section~\ref{Spectroscopy}. 

In Section~\ref{data} we describe the \textit{TESS} photometric data, and the reduction of the {\sc {\textsc{Minerva}}}-Australis spectroscopic data and the radial velocity pipeline, as well as radial velocities collected with other instruments. Section~\ref{analysis} presents the analysis of the data, including the characterization of the host star, the derived properties of the planet, and the limits on any additional planets in the system. In Section~\ref{discussion} we compare TOI-257b with the demographics of the known exoplanets, and discuss the significance of the system. We provide concluding remarks and suggestions for future work in Section~\ref{conclusions}.

\section{Observations and Data Reduction}
\label{data}

TOI-257 (HD 19916) is a bright ($\mathrm{V}=7.612$\,mag) late F-type star, located at a distance of $77.1\pm0.2$\,pc (parallax of $12.9746 \pm 0.0327$\,mas from \textit{Gaia} DR2, \citealt{2018A&A...616A...1G}). The star is slightly evolved with a radius of $1.888\pm0.033$\rsun, mass of $1.390\pm0.046$\msun, and surface gravity of $\log g=4.030\pm0.011$\,dex, derived from the asteroseismic analysis of the \textit{TESS} photometry in Section~\ref{asteroseismology}. The star has an effective temperature of $6075\pm90$\,K and metallicity of $[$M/H$]=0.19\pm0.10$ derived from the analysis of {\textsc{Minerva}}-Australis spectra in Section~\ref{star_spec} as well as a rotational velocity of $v \sin i=11.3 \pm 0.5$\,\kms\ in Section~\ref{star_rot}. TOI-257 has rotational period of $8.07\pm0.27$\,days based on analysis of the \textit{TESS} photometry in Section~\ref{star_rot}.

\begin{table}
\caption{Stellar Parameters for TOI-257. \textbf{Notes.}--$^{\dagger}$Priors used in the \texttt{EXOFASTv2} global fit. $^{\ddagger}$ Broadband magnitudes used in the \texttt{EXOFASTv2} Spectral Energy Distribution analysis. $^{\star}$Upper limit on the V-band extinction from Schegel Dust maps.}
\begin{tabular}{lcc}
\hline
\hline
Parameter & Value & Source \\
\hline
  R.A. (hh:mm:ss)                & 03:10:03.982               & \textit{Gaia} DR2 \\
  Decl. (dd:mm:ss)               & -50:49:56.58                & \textit{Gaia} DR2 \\
  $\mu_{\alpha}$ (mas~yr$^{-1}$) & $97.912 \pm 0.052$         & \textit{Gaia} DR2 \\
  $\mu_{\delta}$ (mas~yr$^{-1}$) & $27.911 \pm 0.082$         & \textit{Gaia} DR2 \\
  Parallax (mas)                 & $12.9746 \pm 0.0327$       & \textit{Gaia} DR2 \\
  $A_{V}$ (mag)                  & $0.0165$ $(\leq0.0506)$\,$^{\dagger,\star}$      & Schegel Dust maps \\
  \vspace{-4pt}
  \\\multicolumn{3}{l}{Broadband Magnitudes:} \vspace{4pt} \\
  $B_{T}$ (mag)                  & $8.293^{+0.020}_{-0.016}$\,$^{\ddagger}$   & Tycho \\
  $V_{T}$ (mag)                  & $7.612^{+0.020}_{-0.011}$\,$^{\ddagger}$   & Tycho \\
  $TESS$ (mag)                   & $7.012 \pm 0.017$          & \textit{TESS} TIC v6 \\
  $J$ (mag)                      & $6.504 \pm 0.020$\,$^{\ddagger}$          & 2MASS \\
  $H$ (mag)                      & $6.325 \pm 0.020$\,$^{\ddagger}$          & 2MASS \\
  $K_{s}$ (mag)                  & $6.256 \pm 0.020$\,$^{\ddagger}$          & 2MASS \\
  $WISE1$ (mag)                  & $6.209 \pm 0.100$\,$^{\ddagger}$          & WISE \\
  $WISE2$ (mag)                  & $6.084 \pm 0.033$\,$^{\ddagger}$          & WISE \\
  $WISE3$ (mag)                  & $6.239^{+0.030}_{-0.015}$\,$^{\ddagger}$   & WISE \\
  $WISE4$ (mag)                  & $6.172^{+0.100}_{-0.048}$\,$^{\ddagger}$   & WISE \\
  $Gaia$ (mag)                   & $7.417^{+0.020}_{-0.000}$\,$^{\ddagger}$   & \textit{Gaia} DR2 \\
  $Gaia_{BP}$ (mag)              & $7.730^{+0.020}_{-0.002}$\,$^{\ddagger}$   & \textit{Gaia} DR2 \\
  $Gaia_{RP}$ (mag)              & $6.994^{+0.020}_{-0.002}$\,$^{\ddagger}$   & \textit{Gaia} DR2 \\
  \vspace{-4pt}
  \\\multicolumn{3}{l}{Spectroscopic Properties from {\sc {\textsc{Minerva}}}-Australis spectra (preferred solution):} \vspace{4pt} \\
  $T_\mathrm{eff}$ (K)           & $6075 \pm 90$\,$^{\dagger}$         & iSpec; this paper \\
  $\log g$ (dex)                 & $3.97 \pm 0.10$        & iSpec; this paper \\
  $[$M/H$]$ (dex)               & $0.19 \pm 0.10$\,$^{\dagger}$          & iSpec; this paper \\
  $R_\star$ ($R_\odot$)          & $1.926 \pm 0.017$      & isochrones; this paper \\
  $M_\star$ ($M_\odot$)          & $1.389^{+0.056}_{-0.009}$        & isochrones; this paper \\
  $\rho_\star$ (g~cm$^{-3}$)     & $0.275 \pm 0.011$        & isochrones; this paper \\
  $L_\star$ ($L_\odot$)          & $4.527 \pm 0.120$       & isochrones; this paper \\
  Age (Gyr)                     & $3.11 \pm 0.46$        & isochrones; this paper \\
  $v \sin i$ (km\,s$^{-1}$)      & $11.3 \pm 0.5$         & LSD; this paper \\
  \vspace{-4pt}
  \\\multicolumn{3}{l}{Spectroscopic Properties from HARPS spectra:} \vspace{4pt} \\
  $T_\mathrm{eff}$ (K)           & $6178 \pm 80$         & ZASPE; this paper \\
  $\log g$ (dex)                 & $4.06 \pm 0.11$        & ZASPE; this paper \\
  $[$Fe/H$]$ (dex)               & $0.32 \pm 0.05$          & ZASPE; this paper \\
    $v \sin i$ (km\,s$^{-1}$)      & $10.2 \pm 0.5$         & ZASPE; this paper \\
\hline
\label{stellar}
\end{tabular}
\end{table}

\subsection{\textit{TESS} Photometry}
\label{TESS}
The star TOI-257 (HD 19916, TIC 200723869 \citealt{2019AJ....158..138S}) was observed in Sectors 3 and 4 by Camera 3 of the \textit{TESS} spacecraft in 2-minute cadence mode nearly continuously between 2018 September 22 and 2018 November 15. The photometric data were processed by the Science Processing Operations Center (SPOC) pipeline as described in \citet{jenkins2016}. Overall, three transits were detected with depth of $\sim1500$\,parts per million (ppm) and duration of $\sim6$\,hours. Two transits are detected in Sector 3 (on BJD\,2458386 and BJD\,2458404), and one in Sector 4 (on BJD\,2458422). The transit at the beginning of Sector 3 was observed during an experiment to improve the spacecraft pointing\footnote{See the data release notes at \url{https://archive.stsci.edu/missions/tess/doc/tess_drn/tess_sector_03_drn04_v02.pdf}}, and the transit in Sector 4 was observed during the thermal ramp.

\begin{figure*}
  \includegraphics[width=\linewidth]{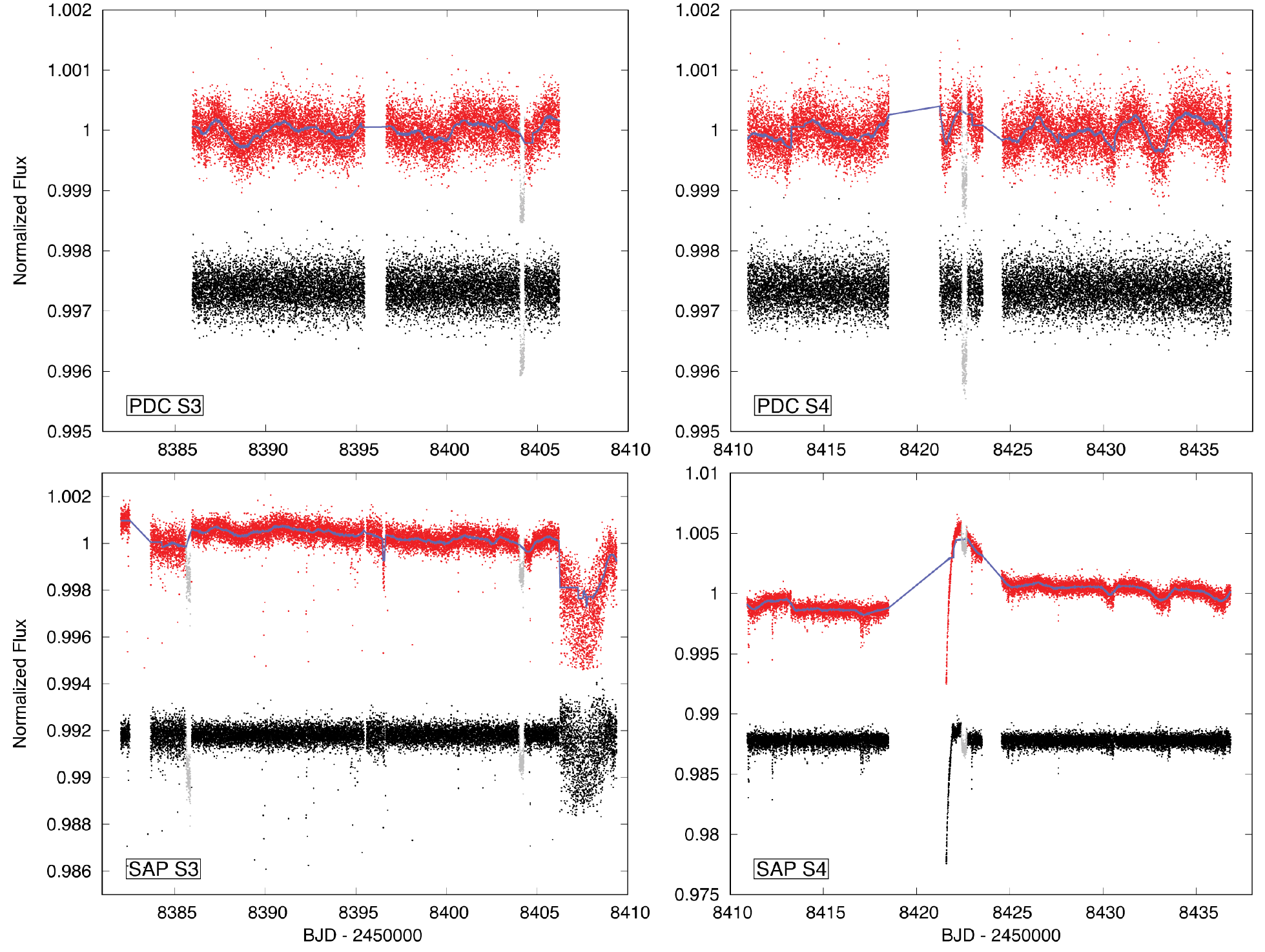}
  \caption{\textit{TESS} light curves of TOI-257 from Sector 3 (left panel) and Sector 4 (right panel). The Pre-search Data Conditioning (PDC, upper panels) and Simple Aperture Photometry (SAP, lower panels) versions of the light curves before (shown in red) and after detrending (shown in black and shifted down arbitrarily to avoid overlap with the red points). The detrending function is blue and transits are grey. Top left: A single transit event was recovered by PDC in Sector 3. Top right: A single transit event was recovered by PDC in Sector 4. Bottom left: Two transit events were recovered by SAP from Sector 3. Bottom right: A single transit event was recovered by SAP in Sector 4.}
  \label{tessphotometry}
\end{figure*}

The \textit{TESS} light curves were accessed from the NASA's Mikulski Archive for Space Telescopes (MAST). The light curves had been processed by the \textit{TESS} team using two different techniques: Pre-search Data Conditioning (PDC, the usual way of light curve extraction and removal of systematics, see, \citealt{jenkins2016}) and Simple Aperture Photometry \citep[SAP, see,][]{2010SPIE.7740E..23T}. These raw SAP and PDC light curves are shown in Figure~\ref{tessphotometry}, along with their detrended versions.

To detrend the PDC light curves, we removed all quality-flagged data (except for stray light flag 2048), clipped $5\sigma$ outliers, removed stellar and instrumental variability, normalized with the mean of the out-of-transit flux, and merged together Sectors 3 and 4. To remove the photometric variability, we used a Savitzky-Golay (SG) filter with a kernel width of 501 data points and a polynomial of order 2 over 3 iterations. During detrending, the planetary transits were masked and then detrended by dividing out the interpolated SG-filtered flux values from the out-of-transit data points. The SG detrending removed any longer-period stellar variability and systematics, and retained any features that occurred on timescales comparable or shorter than the duration of planetary transits \citep{kinemuchi2012,jenkins2016}.

Two transits were recovered using the PDC technique, one in Sector 3 and one in Sector 4. The transit at the beginning of Sector 3 was missed by the PDC procedure since it falls on the part of the light curve that was quality-flagged for manual exclusion during a spacecraft pointing improvement experiment. To recover the first transit in Sector 3, we performed the exact same detrending procedure on the SAP version of the light curve as on the PDC light curve, the only difference being that the manual exclusion (flag 128) data points were not removed. The resulting detrended SAP light curve was used for recovering the first transit observed by \textit{TESS} in Sector 3 but this version of the light curve was not used in the global fit analysis as systematics were not removable as seen in Figure~\ref{tessphotometry}. Indeed the noise level is higher (by $\sim200$\,ppm) in the detrended SAP light curve just prior to this transit event.

\begin{figure}
  \includegraphics[width=\linewidth]{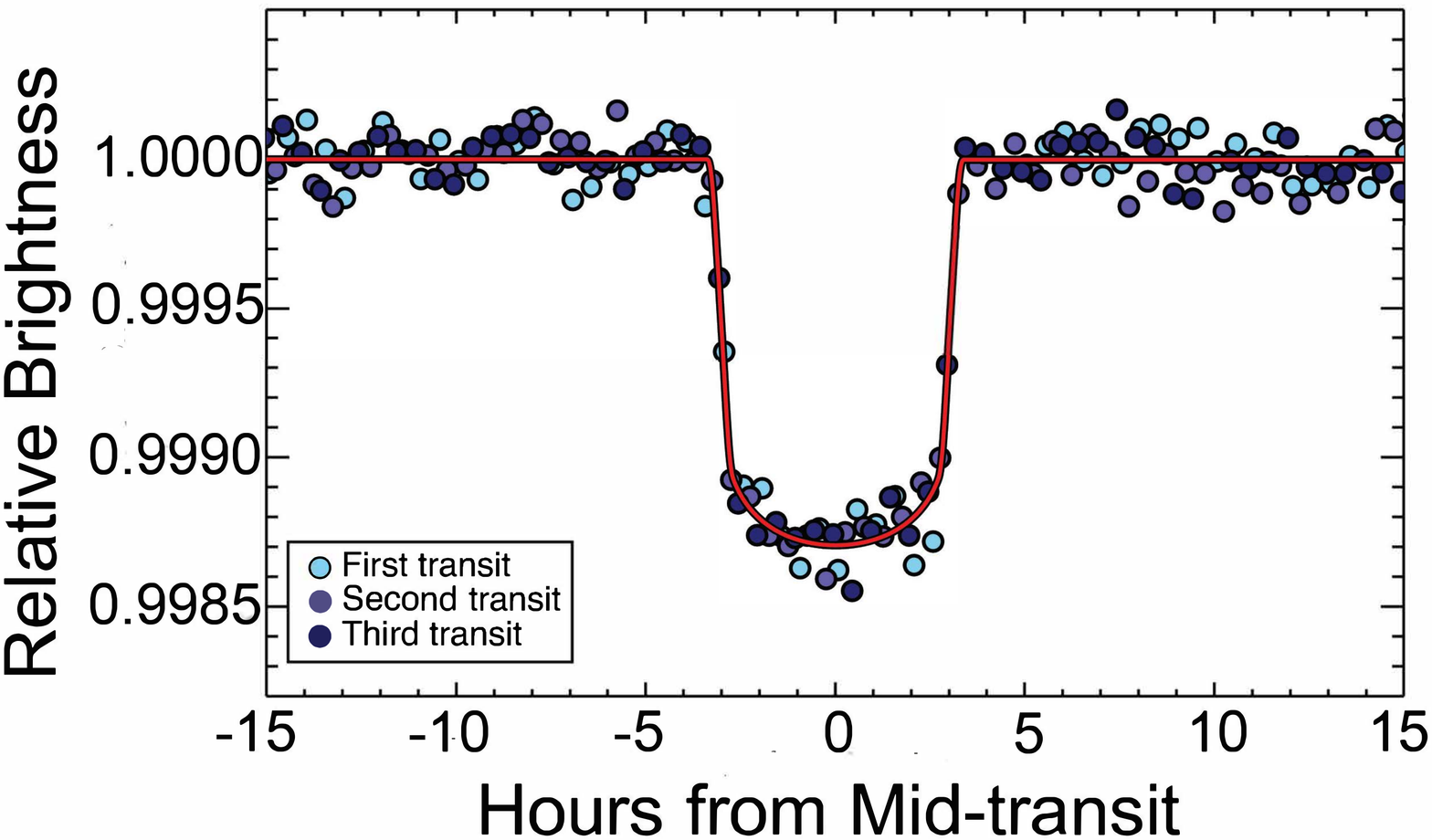}
  \caption{Phase folded \textit{TESS} light curve of TOI-257 binned at a cadence of 30\,minutes with the individual transits color coded showing that they are of similar depth. The first transit comes from the custom light curve where we removed systematics that are the result of a spacecraft pointing anomaly. The second and third transit are from the Pre-search Data Conditioning light curve. The red curve is the best-fit transit model.}
  \label{tesstransits}
\end{figure}

To include the first transit from Sector 3 in the global fit analysis, we created a custom light curve following the procedures of \citet{2019ApJ...881L..19V} to obtain a cleaner light curve relatively free from systematics and stellar variability. We started by using a larger 4.5 pixel radius aperture to extract the Sector 3 photometry, which reduced the amplitude of the systematics observed in the early part of the light curve compared to the \textit{TESS} pipeline's SAP light curve. We then removed systematics from a small segment of the light curve surrounding the first transit (BJD\,$2458383.7 < t < 2458388.0$) by decorrelating against the median background flux value from outside the aperture for each 2-minute image and the standard deviation of the Q1, Q2, and Q3 quaternions within each 2-minute exposure. We excluded points during the planet transit in our decorrelation to prevent the systematics correction from biasing or distorting the shape of the transit. Next, we simultaneously fit the low-frequency variability (which we modeled as a basis spline) with a transit model in a similar manner to \citet{2016ApJS..222...14V}, except that we did not also simultaneously fit for the systematics and we introduced a discontinuity at BJD\,2458385.95 where we switch from the custom light curve to the PDC light curve. The combination of our custom light curve and the PDC light curve are what we use in the final global fitting analysis with \texttt{EXOFASTv2} \citep{2013PASP..125...83E,2017ascl.soft10003E,2019arXiv190709480E}. Figure~\ref{tesstransits} is the resulting 30\,minute binned and phase folded custom light curve along with the PDC light curve and the individual transits color coded.

\subsection{Direct Imaging Follow-up}
\label{DirectImaging}
If a target star has a close companion, the additional flux from the second source can cause photometric contamination, resulting in an underestimated planetary radius, or be the source of an astrophysical false positive. To rule out the presence of close companions, speckle imaging observations were taken of TOI-257 with the SOAR and Zorro instruments.

\subsubsection{SOAR Speckle Imaging}
\label{SOAR}
\begin{figure}
  \includegraphics[width=8.5cm]{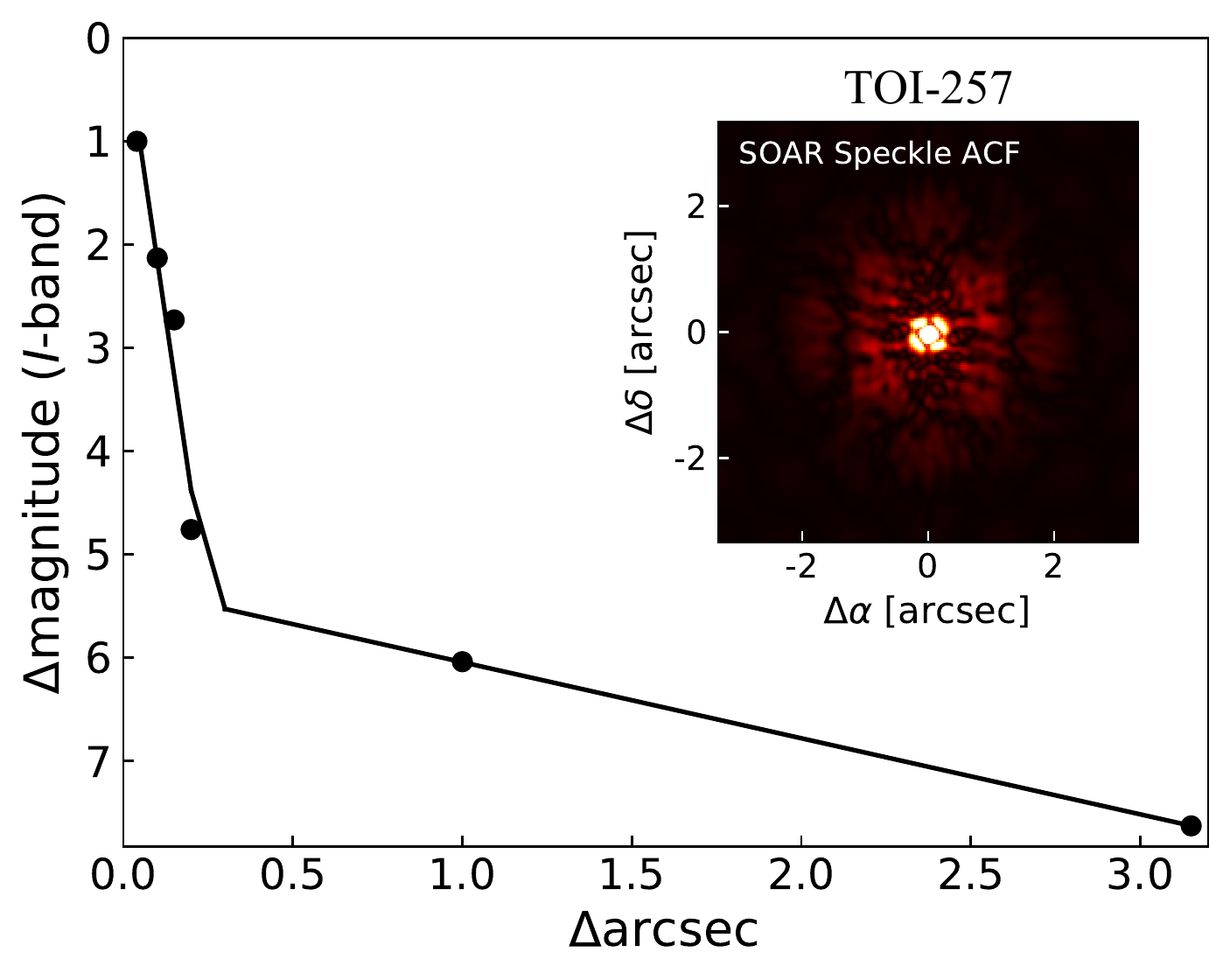}
  \caption{The $5\sigma$ detection sensitivity and inset speckle auto-correlation function from SOAR speckle observing of TOI-257 on 18th February 2019 UT in $I$-band, which is similar to the \textit{TESS} bandpass. The orientation of the inset image has North pointed up and East to the left. No stars were detected within $3\arcsec$ of TOI-257.}
  \label{fig:speckle}
\end{figure}

TOI-257 was observed with SOAR speckle imaging \citep{tokovinin:2018} on 18th February 2019 UT, observing in a similar visible bandpass as \textit{TESS}. The $5\sigma$ detection sensitivity and the speckle auto-correlation function from the SOAR observation are plotted in Figure~\ref{fig:speckle}. Further details of the observations are available in \citet{2020AJ....159...19Z}. No nearby stars were detected within $3\arcsec$ of TOI-257.

\subsubsection{Gemini-South High-Resolution Speckle Imaging using Zorro}
\label{Zorro}
\begin{figure}
  \includegraphics[width=8.0cm]{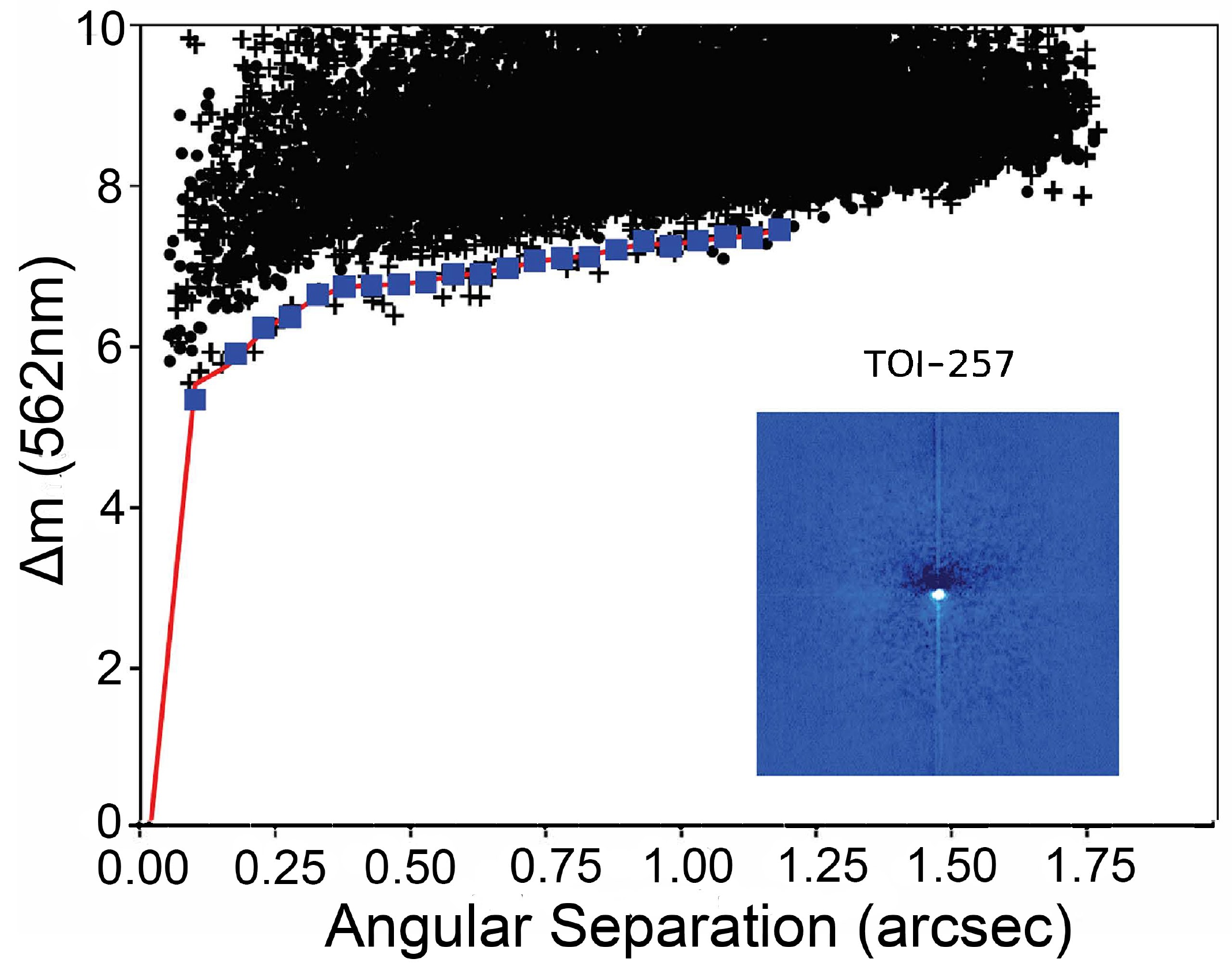}
  \caption{Zorro speckle observation of TOI-257 taken at 562\,nm. Our simultaneous 832\,nm observation provides a similar result. The red line fit and blue points represent the $5\sigma$ fit to the sky level (black points) revealing that no companion star is detected from the diffraction limit (17\,mas) out to 1.75\,\arcsec within a $\Delta$\,mag of 6 to 8. The inset reconstructed speckle image has north up and East to the left and is 2.5\,\arcsec across.}
  \label{fig:zspeckle}
\end{figure}

Direct imaging observations of TOI-257 was also carried out on 12th September 2019 UT using the Zorro speckle instrument on Gemini-South\footnote{https://www.gemini.edu/sciops/instruments/alopeke-zorro/}. Zorro simultaneously provides speckle imaging in two bands, 562\,nm and 832\,nm, with output data products including a reconstructed image, and robust limits on companion detections \citep{howell2011}. Figure~\ref{fig:zspeckle} shows our 562\,nm result and reconstructed speckle image and we find that TOI-257 is indeed a single star with no companion brighter than about 6 magnitudes detected within 1.75\,\arcsec. This limit corresponds to approximately an M3V star at the inner working angle of $\sim0.25$\,\arcsec and M5V at the outer working angle of $\sim1.75$\,\arcsec.

\subsection{Spectroscopy}
\label{Spectroscopy}
We obtained high-resolution spectroscopic observations of TOI-257 with {\sc {\textsc{Minerva}}}-Australis, FEROS, and HARPS to confirm and measure the mass of the \textit{TESS} transiting planet candidate. Here we describe the observations from each spectrograph and list the derived radial velocities in Table~\ref{vels}. 

\subsubsection{High-Resolution Spectroscopy with {\sc {\textsc{Minerva}}}-Australis}
\label{Spectroscopy_Minerva}
%We carried out an intensive radial velocity follow-up campaign with the {\sc {\textsc{Minerva}}}-Australis facility to confirm the planetary nature of the transit-like signals in the \textit{TESS} photometry, measure the mass and orbital properties of the planet, search for any additional planets in the system, and measure the stellar atmospheric properties of the host star.
The {\sc {\textsc{Minerva}}}-Australis facility is an array of five independently operated 0.7\,m CDK700 telescopes located at the Mount Kent Observatory in Queensland, Australia \citep[see,][for a detailed description of the facility]{addison2019}. Designed as a robotic observatory, instruments are remotely accessible and can be operated both in manual or automatic configurations. Four of the telescopes in the array (T2, T3, T4, T6) simultaneously feed stellar light to a single KiwiSpec R4-100 high-resolution spectrograph via fiber optic cables. Only three out of the four telescopes, T3, T4, and T6, were used for spectroscopic observations of TOI-257.

A total of 53 spectra (observations taken simultaneously from multiple telescopes in the array are counted as one observation) of TOI-257 were obtained at 28 epochs between 2019 July 12 and October 15. Each of the telescopes in the {\sc {\textsc{Minerva}}}-Australis array simultaneously feed via 50\,\mm\, circular fiber cables a single KiwiSpec R4-100 high-resolution ($R=80,000$) spectrograph \citep{2012SPIE.8446E..88B} with wavelength coverage from 480 to 630\,nm.

Radial velocities are derived for each telescope using the least-squares technique of \citet{anglada2012} and corrected for spectrograph drifts with simultaneous Thorium-Argon (ThAr) arc lamp observations. We observed TOI-257 with up to three telescopes simultaneously with one or two 20 to 30-minute exposures per epoch, resulting in a signal-to-noise ratio between 30 and 80 per resolution element at $\sim550$\,nm.

The radial velocities from each telescope are given in Table~\ref{vels} labeled by their fiber number. Each telescope (fiber) has its own velocity zero-point which is modeled as a free parameter, and the mean internal uncertainty estimate of the {\sc {\textsc{Minerva}}}-Australis observations is 4.6\mos. The radial velocities collected by {\sc {\textsc{Minerva}}}-Australis show a $\sim10$\,\mos\, sinusoidal variation (RMS uncertainty of 13.9\mos\, based on the residuals from the \texttt{EXOFASTv2} 1-planet fit) that is in phase with the photometric ephemeris with an amplitude compatible with a sub-Saturn-sized planet on a circular orbit as shown in Figures~\ref{rv_full_fig} and \ref{rv_phase_fig}. Additionally, we measured the bisector velocity span (BVS) values using the cross-correlation functions (CCFs) as a check to ensure that the radial velocity variation observed is not from stellar activity or a background eclipsing binary system. As shown in Figure~\ref{bvs_fig}, no correlations are apparent in the BVS values.

\subsubsection{High-Resolution Spectroscopy with the Fiber-fed Extended Range Optical Spectrograph (FEROS)}
\label{Spectroscopy_FEROS}
TOI-257 was observed with the FEROS instrument \citep[$R=48,000$,][]{1999Msngr..95....8K} on the MPG 2.2\,m telescope at La Silla Observatory between 15th December 2018 and 22nd January 2019. We collected a total of eight spectra and the observations were performed in simultaneous calibration mode, utilizing the ThAr arc lamp on the secondary fiber to track and remove instrumental variations due to changes in the temperature and pressure during the science exposures. The exposure times were set to 300\,s, resulting in signal-to-noise ratio between 270 and 370 per resolution element at $\sim510$\,nm. We produced radial velocities by cross-correlation with a G2-type binary mask template using the CERES pipeline \citep{2017PASP..129c4002B}, which also corrects the radial velocities for instrumental systematics and the Earth's motion. 

\subsubsection{High-Resolution Spectroscopy with the High Accuracy Radial velocity Planet Searcher (HARPS)}
\label{Spectroscopy_HARPS}
We monitored TOI-257 with the HARPS spectrograph \citep[$R=$\,120,000,][]{2003Msngr.114...20M} on the ESO 3.6\,m telescope at La Silla Observatory between December 2018 and November 2019. A total of 33 observations were obtained and the data was processed using the CERES pipeline \citep{2017PASP..129c4002B}. The exposure times were set to 300\,s and taken using the high-precision radial velocity mode with simultaneous ThAr, providing a signal-to-noise ratio between 90 and 180 per resolution element at $\sim510$\,nm. We produced radial velocities by cross-correlation with a G2-type binary mask template and derived the stellar properties as $T_\mathrm{eff}=6178 \pm 80$\,K, $\log{g}=4.06 \pm 0.11$\,dex, $[$Fe/H$]=0.32 \pm 0.05$\,dex, and $v\sin i=10.2 \pm 0.5$\,km\,s$^{-1}$\, for the host star with the HARPS spectra using ZASPE \citep{2017MNRAS.467..971B}. The metallicity results from the HARPS spectra indicate that the star is definitively metal rich.

\begin{table}
\caption{Journal of radial velocity observations of TOI-257. \textbf{Notes.}--M-A Tel3, M-A Tel4, and M-A Tel6 are {\textsc{Minerva}}-Australis Telescope3, Telescope4, and Telescope5, respectively.}
\begin{tabular}{cccc}
\hline
\hline
Date & RV & $\sigma$ & Instrument \\
(BJD) & (m\,s$^{-1}$) & (m\,s$^{-1}$) & \\
\hline
        2458465.539980	&	21.9	&	2.0	&	HARPS		\\
	    2458465.602650	&	26.7	&	2.0	&	HARPS		\\
	    2458465.690670	&	26.1	&	2.0	&	HARPS		\\
	    2458466.529660	&	24.8	&	2.0	&	HARPS		\\
	    2458466.591590	&	17.1	&	2.0	&	HARPS		\\
	    2458466.678080	&	23.5	&	2.0	&	HARPS		\\
	    2458466.682320	&	22.5	&	2.0	&	HARPS		\\
	    2458467.674470	&	12.5	&	5.3	&	FEROS		\\
	    2458468.663190	&	8.1	    &	5.5	&	FEROS		\\
	    2458481.588670	&	20.9	&	2.0	&	HARPS		\\
	    2458481.593290	&	24.6	&	2.0	&	HARPS		\\
	    2458481.597630	&	24.1	&	2.0	&	HARPS		\\
	    2458482.673800	&	32.1	&	2.0	&	HARPS		\\
	    2458482.678140	&	29.7	&	2.0	&	HARPS		\\
	    2458493.714430	&	-11.4	&	6.2	&	FEROS		\\
	    2458497.608960	&	-10.3	&	5.7	&	FEROS		\\
	    2458500.629830	&	-19.3	&	5.7	&	FEROS		\\
	    2458505.566740	&	-14.6	&	5.9	&	FEROS		\\
	    2458677.272975	&	10.8	&	3.0	&	M-A	Tel3	\\
	    2458677.272975	&	-13.2	&	3.4	&	M-A	Tel4	\\
	    2458677.294387	&	24.3	&	3.1	&	M-A	Tel3	\\
	    2458677.294387	&	10.3	&	3.4	&	M-A	Tel4	\\
	    2458680.203692	&	11.6	&	3.6	&	M-A	Tel3	\\
	    2458680.203692	&	20.7	&	4.1	&	M-A	Tel4	\\
	    2458680.203692	&	-8.2	&	7.5	&	M-A	Tel6	\\
	    2458680.225093	&	0.1	    &	3.9	&	M-A	Tel3	\\
	    2458680.225093	&	3.3	    &	3.8	&	M-A	Tel4	\\
	    2458680.225093	&	5.4	    &	8.0	&	M-A	Tel6	\\
	    2458681.170185	&	1.9	    &	3.5	&	M-A	Tel3	\\
	    2458681.170185	&	22.1	&	3.6	&	M-A	Tel4	\\
	    2458681.170185	&	-8.3	&	4.6	&	M-A	Tel6	\\
	    2458681.191597	&	-3.8	&	3.3	&	M-A	Tel3	\\
	    2458681.191597	&	-11.9	&	3.9	&	M-A	Tel4	\\
	    2458681.191597	&	14.9	&	4.6	&	M-A	Tel6	\\
	    2458682.146655	&	25.4	&	3.9	&	M-A	Tel3	\\
	    2458682.146655	&	27.6	&	7.2	&	M-A	Tel4	\\
	    2458682.146655	&	12.7	&	5.3	&	M-A	Tel6	\\
	    2458682.168067	&	14.9	&	3.9	&	M-A	Tel3	\\
	    2458682.168067	&	19.6	&	4.6	&	M-A	Tel4	\\
	    2458682.168067	&	5.0	    &	5.6	&	M-A	Tel6	\\
	    2458683.249780	&	6.3	    &	3.5	&	M-A	Tel4	\\
	    2458683.276111	&	-5.9	&	4.6	&	M-A	Tel3	\\
	    2458683.276111	&	14.6	&	3.3	&	M-A	Tel4	\\
        2458688.201840	&	5.1	&	3.0	&	M-A	Tel3	\\
        2458688.201840	&	-0.1	&	3.3	&	M-A	Tel4	\\
        2458688.201840	&	-22.9	&	6.0	&	M-A	Tel6	\\
        2458688.223252	&	-23.0	&	2.8	&	M-A	Tel3	\\
        2458688.223252	&	-5.6	&	3.1	&	M-A	Tel4	\\
        2458688.223252	&	-25.9	&	6.5	&	M-A	Tel6	\\
        2458689.179745	&	-2.6	&	3.5	&	M-A	Tel3	\\
        2458689.179745	&	1.9	&	4.1	&	M-A	Tel4	\\
        2458689.179745	&	4.6	&	5.1	&	M-A	Tel6	\\
        2458689.201146	&	0.1	&	3.1	&	M-A	Tel3	\\
        2458689.201146	&	-0.9	&	3.5	&	M-A	Tel4	\\
        2458689.201146	&	13.6	&	4.4	&	M-A	Tel6	\\
        2458694.193565	&	17.4	&	2.6	&	M-A	Tel3	\\
        2458694.193565	&	22.3	&	2.9	&	M-A	Tel4	\\
        2458694.193565	&	28.4	&	3.8	&	M-A	Tel6	\\
        2458694.214965	&	8.0	&	2.5	&	M-A	Tel3	\\
        2458694.214965	&	3.3	&	2.7	&	M-A	Tel4	\\
\hline
\label{vels}
\end{tabular}
\end{table}

\begin{table}
\contcaption{}
\begin{tabular}{cccc}
\hline
\hline
Date & RV & $\sigma$ & Instrument \\
(BJD) & (m\,s$^{-1}$) & (m\,s$^{-1}$) & \\
\hline
        2458694.214965	&	9.7	&	3.8	&	M-A	Tel6	\\
        2458695.195069	&	7.9	&	2.8	&	M-A	Tel3	\\
        2458695.195069	&	16.6	&	3.2	&	M-A	Tel4	\\
        2458695.195069	&	26.1	&	3.7	&	M-A	Tel6	\\
        2458708.118403	&	20.7	&	3.6	&	M-A	Tel3	\\
        2458708.118403	&	10.3	&	4.7	&	M-A	Tel4	\\
        2458708.118403	&	7.4	&	5.5	&	M-A	Tel6	\\
        2458708.139815	&	15.4	&	3.3	&	M-A	Tel3	\\
        2458708.139815	&	-0.5	&	4.0	&	M-A	Tel4	\\
        2458708.139815	&	4.4	&	4.9	&	M-A	Tel6	\\
        2458710.171273	&	3.3	&	3.7	&	M-A	Tel3	\\
        2458710.171273	&	21.9	&	3.7	&	M-A	Tel4	\\
        2458710.171273	&	-9.4	&	4.6	&	M-A	Tel6	\\
        2458710.192674	&	13.6	&	3.5	&	M-A	Tel3	\\
        2458710.192674	&	14.9	&	3.4	&	M-A	Tel4	\\
        2458710.192674	&	13.0	&	4.4	&	M-A	Tel6	\\
        2458712.165914	&	34.1	&	3.7	&	M-A	Tel3	\\
        2458712.165914	&	12.5	&	3.9	&	M-A	Tel4	\\
        2458712.165914	&	10.9	&	4.6	&	M-A	Tel6	\\
        2458712.187326	&	34.5	&	3.4	&	M-A	Tel3	\\
        2458712.187326	&	12.9	&	3.7	&	M-A	Tel4	\\
        2458712.187326	&	20.1	&	4.3	&	M-A	Tel6	\\
        2458714.125567	&	4.3	&	5.0	&	M-A	Tel3	\\
        2458714.125567	&	-0.6	&	5.6	&	M-A	Tel4	\\
        2458714.125567	&	-7.3	&	6.4	&	M-A	Tel6	\\
        2458714.146979	&	27.6	&	4.5	&	M-A	Tel3	\\
        2458714.146979	&	-9.1	&	4.8	&	M-A	Tel4	\\
        2458714.146979	&	6.2	&	5.6	&	M-A	Tel6	\\
        2458715.217465	&	12.2	&	2.7	&	M-A	Tel3	\\
        2458715.217465	&	15.8	&	3.8	&	M-A	Tel4	\\
        2458715.217465	&	-8.1	&	4.2	&	M-A	Tel6	\\
        2458715.238877	&	6.6	&	2.6	&	M-A	Tel3	\\
        2458715.238877	&	-4.4	&	3.5	&	M-A	Tel4	\\
        2458715.238877	&	6.2	&	4.1	&	M-A	Tel6	\\
        2458716.222002	&	-15.2	&	2.6	&	M-A	Tel3	\\
        2458716.222002	&	7.3	&	4.0	&	M-A	Tel4	\\
        2458716.222002	&	1.9	&	4.3	&	M-A	Tel6	\\
        2458716.243403	&	-15.5	&	2.6	&	M-A	Tel3	\\
        2458716.243403	&	-0.9	&	4.0	&	M-A	Tel4	\\
        2458716.243403	&	-6.4	&	4.2	&	M-A	Tel6	\\
        2458719.145729	&	-36.6	&	3.4	&	M-A	Tel3	\\
        2458719.145729	&	-38.1	&	3.9	&	M-A	Tel4	\\
        2458719.145729	&	-22.8	&	4.8	&	M-A	Tel6	\\
        2458720.110220	&	-17.3	&	3.1	&	M-A	Tel3	\\
        2458720.110220	&	-24.6	&	3.8	&	M-A	Tel4	\\
        2458720.110220	&	-37.4	&	4.8	&	M-A	Tel6	\\
        2458720.131632	&	-29.8	&	3.0	&	M-A	Tel3	\\
        2458720.131632	&	-24.1	&	3.7	&	M-A	Tel4	\\
        2458720.131632	&	-29.0	&	4.4	&	M-A	Tel6	\\
        2458722.130023	&	-5.8	&	3.3	&	M-A	Tel3	\\
        2458722.130023	&	-16.4	&	5.7	&	M-A	Tel4	\\
        2458722.130023	&	-14.2	&	4.9	&	M-A	Tel6	\\
        2458722.151435	&	-16.2	&	3.2	&	M-A	Tel3	\\
        2458722.151435	&	-8.9	&	5.8	&	M-A	Tel4	\\
        2458722.151435	&	-21.1	&	4.8	&	M-A	Tel6	\\
        2458725.111910	&	-8.4	&	2.9	&	M-A	Tel3	\\
        2458725.111910	&	12.8	&	5.1	&	M-A	Tel4	\\
        2458725.111910	&	-20.3	&	4.6	&	M-A	Tel6	\\
        2458725.133322	&	5.0	&	2.8	&	M-A	Tel3	\\
        2458725.133322	&	-25.7	&	5.1	&	M-A	Tel4	\\
        2458725.133322	&	10.8	&	4.4	&	M-A	Tel6	\\
        2458728.105961	&	-2.4	&	3.1	&	M-A	Tel3	\\
        2458728.105961	&	-4.3	&	4.2	&	M-A	Tel4	\\
\hline
\end{tabular}
\end{table}

\begin{table}
\contcaption{}
\begin{tabular}{cccc}
\hline
\hline
Date & RV & $\sigma$ & Instrument \\
(BJD) & (m\,s$^{-1}$) & (m\,s$^{-1}$) & \\
\hline
        2458728.105961	&	15.3	&	4.3	&	M-A	Tel6	\\
        2458728.127373	&	-1.2	&	3.1	&	M-A	Tel3	\\
        2458728.127373	&	22.4	&	4.0	&	M-A	Tel4	\\
        2458728.127373	&	28.7	&	4.2	&	M-A	Tel6	\\
        2458729.072407	&	-1.2	&	3.9	&	M-A	Tel3	\\
        2458729.072407	&	-10.4	&	4.4	&	M-A	Tel4	\\
        2458729.072407	&	-6.1	&	5.4	&	M-A	Tel6	\\
        2458729.093808	&	-2.2	&	3.7	&	M-A	Tel3	\\
        2458729.093808	&	8.2	&	4.4	&	M-A	Tel4	\\
        2458729.093808	&	4.2	&	5.4	&	M-A	Tel6	\\
        2458729.115220	&	-15.5	&	3.6	&	M-A	Tel3	\\
        2458729.115220	&	5.8	&	4.2	&	M-A	Tel4	\\
        2458729.115220	&	-2.7	&	5.3	&	M-A	Tel6	\\
        2458730.018252	&	-1.8	&	5.2	&	M-A	Tel3	\\
        2458730.018252	&	-2.9	&	7.5	&	M-A	Tel4	\\
        2458730.018252	&	-2.0	&	6.7	&	M-A	Tel6	\\
        2458730.039664	&	25.0	&	4.4	&	M-A	Tel3	\\
        2458730.039664	&	-4.7	&	4.7	&	M-A	Tel4	\\
        2458730.039664	&	-3.9	&	5.9	&	M-A	Tel6	\\
        2458734.108032	&	21.5	&	3.7	&	M-A	Tel3	\\
        2458734.108032	&	7.9	&	4.4	&	M-A	Tel4	\\
        2458734.108032	&	-17.1	&	5.2	&	M-A	Tel6	\\
        2458734.129444	&	-5.7	&	3.5	&	M-A	Tel3	\\
        2458734.129444	&	0.4	&	4.1	&	M-A	Tel4	\\
        2458734.129444	&	33.6	&	4.4	&	M-A	Tel6	\\
        2458735.062465	&	3.5	&	3.9	&	M-A	Tel3	\\
        2458735.062465	&	-22.2	&	4.5	&	M-A	Tel4	\\
        2458735.062465	&	1.7	&	5.3	&	M-A	Tel6	\\
        2458735.083877	&	-1.0	&	3.7	&	M-A	Tel3	\\
        2458735.083877	&	-22.1	&	4.4	&	M-A	Tel4	\\
        2458735.083877	&	11.7	&	5.0	&	M-A	Tel6	\\
        2458737.059757	&	-17.9	&	4.1	&	M-A	Tel3	\\
        2458737.059757	&	-13.1	&	4.7	&	M-A	Tel4	\\
        2458737.059757	&	-28.1	&	4.7	&	M-A	Tel6	\\
        2458737.081169	&	-13.7	&	3.9	&	M-A	Tel3	\\
        2458737.081169	&	-11.8	&	4.4	&	M-A	Tel4	\\
        2458737.081169	&	-6.5	&	4.7	&	M-A	Tel6	\\
        2458739.195799	&	-23.8	&	3.5	&	M-A	Tel3	\\
        2458739.195799	&	5.5	&	5.0	&	M-A	Tel4	\\
        2458739.195799	&	-15.9	&	4.4	&	M-A	Tel6	\\
        2458739.217211	&	-23.8	&	3.7	&	M-A	Tel3	\\
        2458739.217211	&	-15.5	&	4.2	&	M-A	Tel4	\\
        2458739.217211	&	5.5	&	4.9	&	M-A	Tel6	\\
        2458741.221794	&	-3.5	&	3.5	&	M-A	Tel3	\\
        2458741.221794	&	6.5	&	3.7	&	M-A	Tel4	\\
        2458741.221794	&	4.1	&	4.6	&	M-A	Tel6	\\
        2458742.066331	&	19.7	&	5.3	&	M-A	Tel3	\\
        2458742.066331	&	-17.8	&	4.3	&	M-A	Tel4	\\
        2458742.066331	&	-11.7	&	6.2	&	M-A	Tel6	\\
        2458760.781220	&	32.2	&	2.3	&	HARPS		\\
        2458762.784910	&	23.1	&	2.0	&	HARPS		\\
        2458764.688350	&	39.6	&	2.0	&	HARPS		\\
        2458765.692580	&	38.7	&	2.5	&	HARPS		\\
        2458772.692230	&	23.3	&	2.0	&	HARPS		\\
        2458773.680520	&	34.7	&	2.1	&	HARPS		\\
        2458774.741390	&	24.4	&	2.0	&	HARPS		\\
        2458775.824000	&	9.3	&	2.6	&	HARPS		\\
        2458777.807510	&	15.8	&	3.2	&	HARPS		\\
\hline
\end{tabular}
\end{table}

\begin{table}
\contcaption{}
\begin{tabular}{cccc}
\hline
\hline
Date & RV & $\sigma$ & Instrument \\
(BJD) & (m\,s$^{-1}$) & (m\,s$^{-1}$) & \\
\hline
        2458780.869940	&	21.2	&	9.7	&	HARPS		\\
        2458802.637920	&	21.6	&	2.0	&	HARPS		\\
        2458804.705790	&	37.1	&	2.0	&	HARPS		\\
        2458806.677240	&	26.1	&	2.0	&	HARPS		\\
        2458810.664000	&	14.8	&	2.0	&	HARPS		\\
        2458811.725990	&	7.1	&	2.0	&	HARPS		\\
        2458813.686300	&	20.1	&	2.0	&	HARPS		\\
        2458833.676260	&	20.9	&	2.0	&	HARPS		\\
\hline
\end{tabular}
\end{table}

\section{Analysis}
\label{analysis}

\subsection{Host Star Properties from Spectroscopy}
\label{star_spec}
\begin{figure}
  \includegraphics[width=8.5cm]{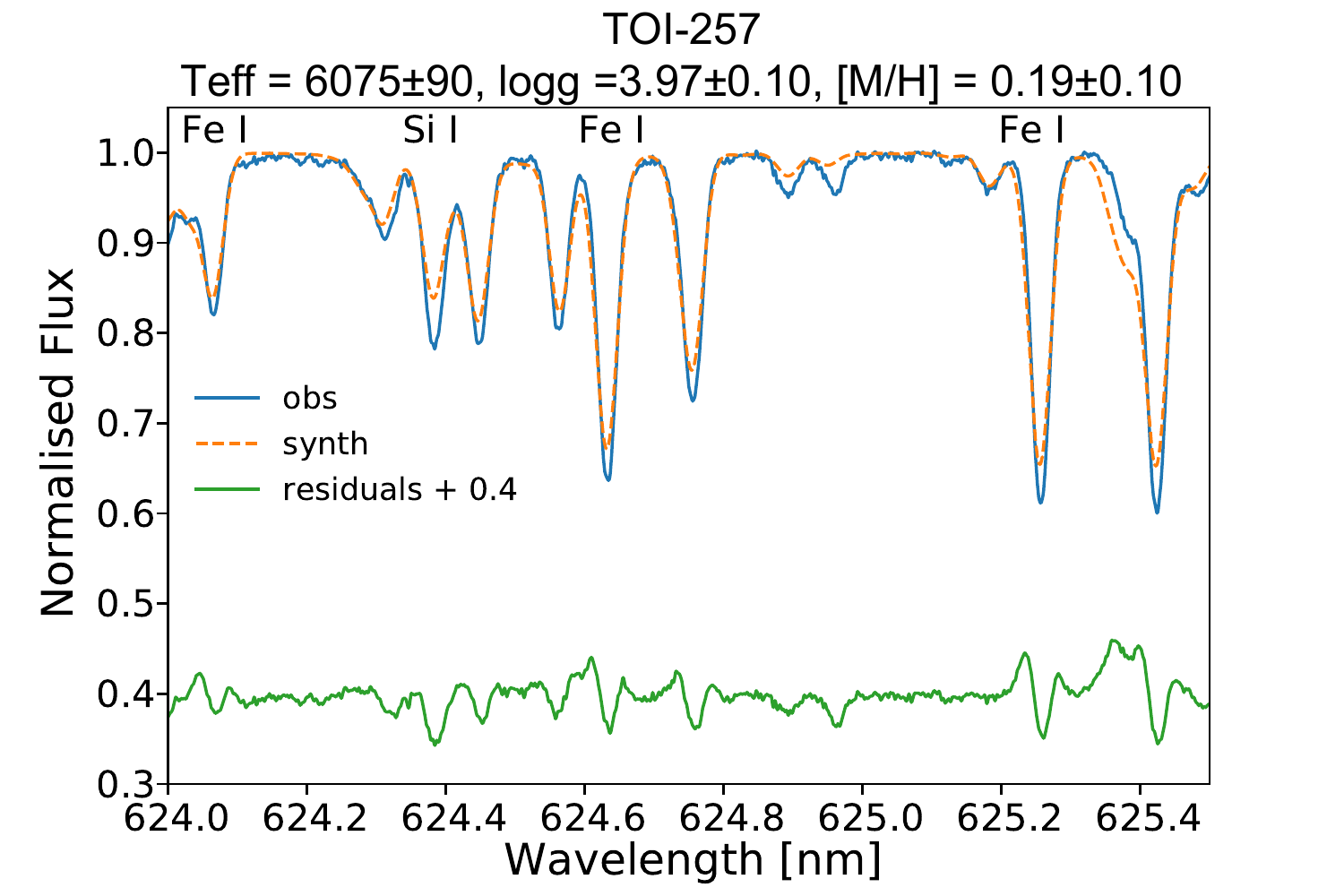}
  \caption{The best fit synthetic model spectrum from \textsc{iSpec} (the red dashed line) of TOI-257 to that of the combined stellar spectrum obtained from the {\sc {\textsc{Minerva}}}-Australis spectroscopic observations (the blue solid line) for the wavelength region between 624.0\,nm and 625.5\,nm. The residuals of the fit are shown as the green solid line.}
  \label{spectrum}
\end{figure}

We used the {\sc {\textsc{Minerva}}}-Australis spectra to determine TOI-257's atmospheric stellar parameters. Through the \textsc{python} package \textsc{iSpec} \citep{ispec2014,ispec2019}, we stacked the stellar spectra to derive the effective temperature, surface gravity, and overall metallicity ([M/H]) of the star. We configured the \textsc{iSpec} synthetic grid to incorporate a MARCS atmospheric model \citep{MARCSmodel} and utilized the \textsc{spectrum} \citep{spectrumcode} radiative transfer code. [M/H] was derived using version 5.0 of {\it Gaia\/}-ESO Survey's (GES) line-list \citep{GESlinelist} normalized by solar values obtained by \citet{Asplund09}. Our synthetic spectra fit was constructed by setting initial values for \teff, $\log{g}$ and [M/H] of 6050\,K, 4.44\,dex, and 0.00\,dex, respectively, based on the parameters from a broadband spectral energy distribution (SED) analysis with \texttt{EXOFASTv2}. Figure \ref{spectrum} depicts our observed spectra and synthetic model produced by \textsc{iSpec}. Our derived \teff, $\log{g}$ and [M/H] values were then fed into the Bayesian isochrone modeler \textsc{isochrones} \citep{isochrones,2015ApJ...809...25M} that uses the Dartmouth Stellar Evolution Database \citep{2008ApJS..178...89D}.

\textsc{isochrones} uses nested sampling scheme called \textsc{multinest} \citep{2009MNRAS.398.1601F} to determine the stellar mass, radius, and age, which was then used to derive the stellar density and luminosity of TOI-257. For this particular analysis, we used the stellar parameter results from \textsc{iSpec} as well as the parallax value from \textit{Gaia} DR2 with G, H, J, K, V and W1 magnitudes\footnote{From experience, we find that \textsc{isochrones} delivers more reliable results when using just the G, H, J, K, V and W1 magnitudes instead of all the available magnitudes given in the literature for a star.} as priors in the global fit. The spectroscopic stellar \textsc{iSpec} and \textsc{isochrones} values can be found in Table \ref{stellar} and are in good agreement with the SED analysis performed using \texttt{EXOFASTv2} and the asteroseismology. We then incorporated the \teff\ and [M/H] values derived from the \textsc{iSpec} analysis of the {\sc {\textsc{Minerva}}}-Australis spectra as priors in the final \texttt{EXOFASTv2} global fit of the data in Section \ref{global_fit} and stellar luminosity of $L=4.57 \pm 0.16$\,\lsun\ derived from SED fitting as a prior in the asteroseismology analysis in Section \ref{asteroseismology}. We also note that stellar atmospheric parameters derived from the HARPS spectra are in general agreement with the ones derived with the {\sc {\textsc{Minerva}}}-Australis spectra, though the HARPS spectra suggest that the star is definitely metal rich ($[$Fe/H$]=0.32 \pm 0.05$\,dex) while the {\sc {\textsc{Minerva}}}-Australis spectra is compatible with solar metallicity to within $2\sigma$ ($[$M/H$]=0.19 \pm 0.10$\,dex). Given that the SED analysis is in better agreement with the derived stellar parameter from the {\sc {\textsc{Minerva}}}-Australis spectra and the strong degeneracies in the model atmospheres with parameters such as \teff, metallicity, and $\log{g}$ \citep[e.g., see,][]{2016ApJS..226....4H}, we have chosen to use the stellar atmospheric parameters from {\sc {\textsc{Minerva}}}-Australis in the global analysis.

\subsection{Asteroseismology}
\label{asteroseismology}

\subsubsection{Global Asteroseismic Parameters}
To perform asteroseismic analysis on TOI-257 we produced a custom light curve using the \textit{TESS} Asteroseismic Science Operations Center \citep[TASOC,][]{lund17} photometry pipeline\footnote{\url{https://tasoc.dk/code/}} (Handberg et al., in prep.), which is based on software originally developed to generate light curves for data collected by the K2 Mission \citep{lund15}. The TASOC pipeline implements a series of corrections to optimize light curves for an asteroseismic analysis \citep{handberg14}, including the removal of instrumental artefacts and of the transit events using a combination of filters utilizing the estimated planetary period. The photometric performance of the TASOC light curve was comparable to the light curve produced by the SPOC pipeline.

\begin{figure}
\begin{center}
\resizebox{\hsize}{!}{\includegraphics{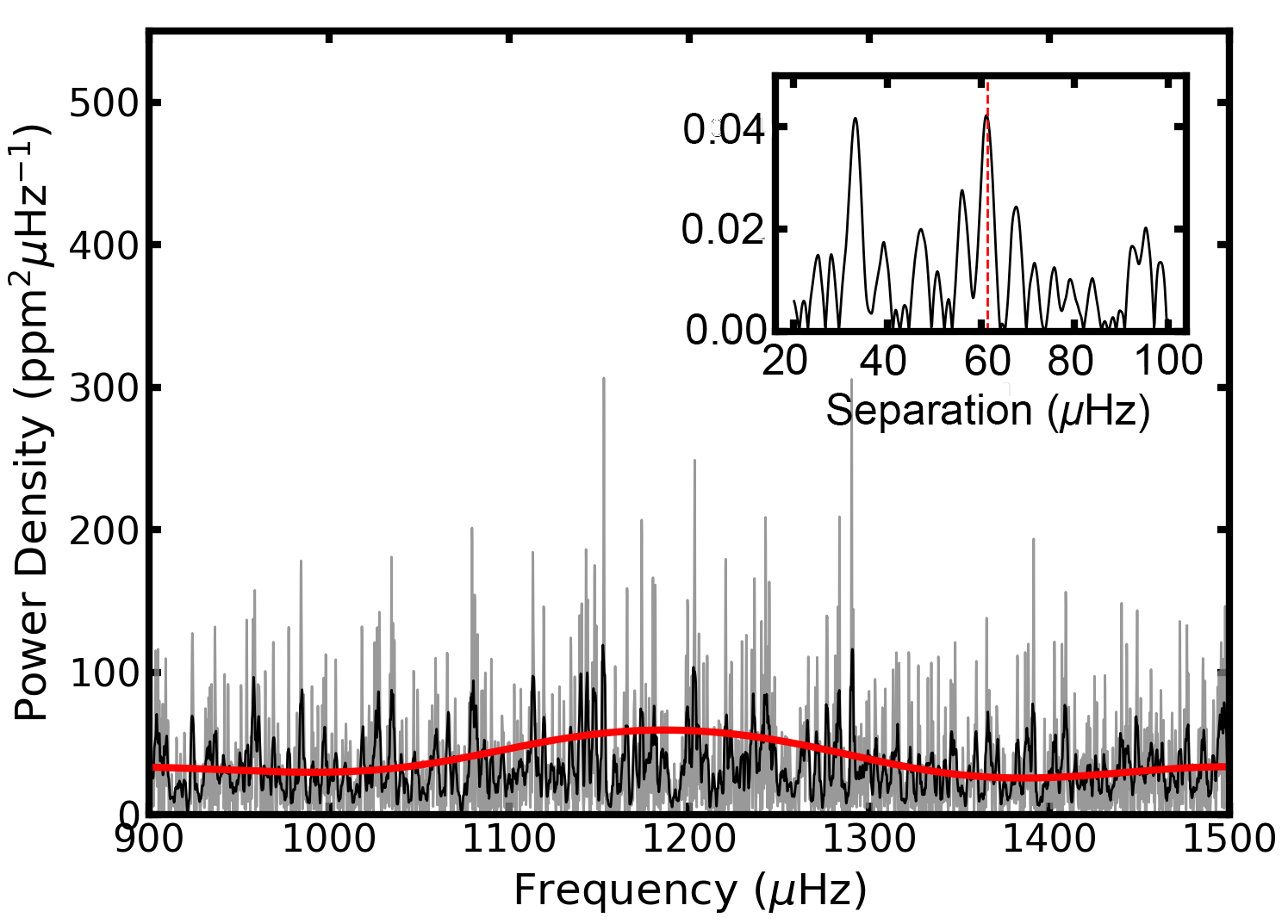}}
\caption{Power spectrum of the Sector 3 TASOC light curve of \target\ (grey line). The black and red lines show the power spectrum smoothed with a boxcar width of $2$\,\muHz\ and Gaussian with a full width at half max of $\dnu=61.4$, respectively. The inset shows the autocorrelation of the power spectrum, with a red line marking the expected value of \dnu\ based on the location of the power excess.}
\label{fig:seismo}
\end{center}
\end{figure}

Solar-like oscillations are broadly described by a frequency of maximum oscillation power (\numax) and a large frequency separation (\dnu), which approximately scale with $\log g$\ and the mean stellar density, respectively \citep[see,][]{2019LRSP...16....4G}. The power spectrum of the Sector 3 light curve of \target\ displays a power excess near $\sim$\,1200\,\muHz\ (Figure~\ref{fig:seismo}), consistent with the spectroscopic $\log g$\ and the expected frequency range from the \textit{TESS} asteroseismic target list \citep[ATL,][]{2019ApJS..241...12S}. An autocorrelation of the power spectrum reveals a peak at a frequency spacing consistent with the location of the excess power \citep[e.g.][]{stello09}. Furthermore, the amplitude of the power excess ($\sim$ 9\,ppm) is consistent with the expected value from observations by \textit{Kepler} \citep{huber11}. The addition of the Sector 4 light curve reduced the significance of the asteroseismic detection due to the slightly elevated noise level, and was thus discarded for the remainder of our analysis.

To test the significance of the detection and measure \numax\ and \dnu\ we used 15 independent analysis methods within working group 1 of the \textit{TESS} Asteroseismic Science Consortium \citep[e.g.][]{huber09,mathur10b,mosser11c,benomar12,kallinger12,corsaro14,2016AN....337..774D,campante2018}. All but one pipeline reported a significant detection of solar-like oscillations. The final parameters are $\numax=1188 \pm 40$\,$\muHz$ and $\dnu = 61.4 \pm 1.5$\,$\muHz$, with the central value taken from the solution closest to the median of all solutions, and uncertainties calculated from the median formal uncertainty returned by individual pipelines added in quadrature to the scatter over individual methods.

\subsubsection{Grid-Based Modeling}
We used a number of independent approaches to model the observed global asteroseismic parameters, including different stellar evolution codes \citep[ASTEC, GARSTEC, MESA, and YREC,][]{jcd08,weiss08,paxton11,paxton13,paxton15,choi16,demarque08} and modeling methods \citep[BeSPP, BASTA, PARAM, isoclassify,][]{silva15,serenelli17,rodrigues14,rodrigues17,huber17,2018MNRAS.476.1470G}.
Model inputs included the spectroscopic temperature and metallicity (see Section~\ref{star_spec}), \numax, \dnu, and the luminosity derived from the Gaia parallax and SED fitting. To investigate the effects of different input parameters, modelers were asked to provide solutions with and without taking into account the luminosity constraint. 

The modeling results showed a bi-modality in mass (and thus age) at $\sim 1.2$\msun\ and $\sim 1.4$\msun, with all pipelines favoring the higher mass solution once the luminosity constraint was included. We adopted the solution closest to the median of all returned values, with uncertainties calculated by adding the median uncertainty for a given stellar parameter in quadrature to the standard deviation of the parameter for all methods. This method has been commonly adopted for \kep\ \citep[e.g.][]{chaplin14} and captures both random and systematic errors estimated from the spread among different methods. The final estimates of the stellar parameters, taking into account the luminosity constraint, are summarized in Table~3, constraining the radius, mass, density and age of \target\ to $\sim$\,2\,\%, $\sim$\,3\,\%, $\sim$\,3\,\% and $\sim$\,13\,\%. We emphasize that these uncertainties in stellar parameters are robust against systematic errors from different stellar model grids, which are frequently neglected when characterizing exoplanets. The stellar mass and radius derived from this analysis is used as priors in the final \texttt{EXOFASTv2} global fit of the data in Section~\ref{global_fit}.

\begin{table}
\begin{center}
\caption{Asteroseismic Stellar Parameters for TOI-257. \textbf{Notes:} $^{\dagger}$Priors used in the \texttt{EXOFASTv2} global fit.}
\renewcommand{\tabcolsep}{0mm}
\begin{tabular}{l c}
\hline
\hline
\noalign{\smallskip}
\multicolumn{2}{c}{Input Parameters} \\
\noalign{\smallskip}
\hline
\noalign{\smallskip}
Frequency of maximum oscillation power, \numax\ (\muHz), & $1188 \pm 40$ \\
Large frequency separation, \dnu\ (\muHz), & $61.4 \pm 1.5$ \\
Effective Temperature, \teff\, (K) & \teffstar \\
Metallicity, [Fe/H] (dex) & \fehstar \\
Luminosity, $L$ (\lsun) & $4.57 \pm 0.16$ \\
\noalign{\smallskip}
\hline
\multicolumn{2}{c}{Stellar Parameters} \\
\noalign{\smallskip}
\hline
\noalign{\smallskip}
Stellar Mass, \mstar\ (\msun)& \massstar\,$^{\dagger}$ \\
Stellar Radius, \rstar\ (\rsun)& \radstar\,$^{\dagger}$ \\
Stellar Density, \rhostar\ (cgs)& \denstar \\
Surface Gravity, $\log g$\ (cgs) & \loggstar \\
Age, $t$ (Gyr) & \agestar \\
\noalign{\smallskip}
\hline
\end{tabular}
\end{center}
\label{seismo}
\end{table}

\subsection{Stellar Rotation Period Estimates}
\label{star_rot}
The rotation period of TOI-257 was derived from the estimated stellar radius and by performing Lomb-Scargle \citep{1982ApJ...263..835S} periodogram and auto-correlation function analysis \citep[e.g.,][]{2013MNRAS.432.1203M} on the \textit{TESS} lightcurve, and measuring the projected stellar rotation velocity ($v\sin i$) from {\sc {\textsc{Minerva}}}-Australis spectra, assuming the axis of stellar rotation is perpendicular to the line of sight. 

\begin{figure*}
  \includegraphics[width=\linewidth]{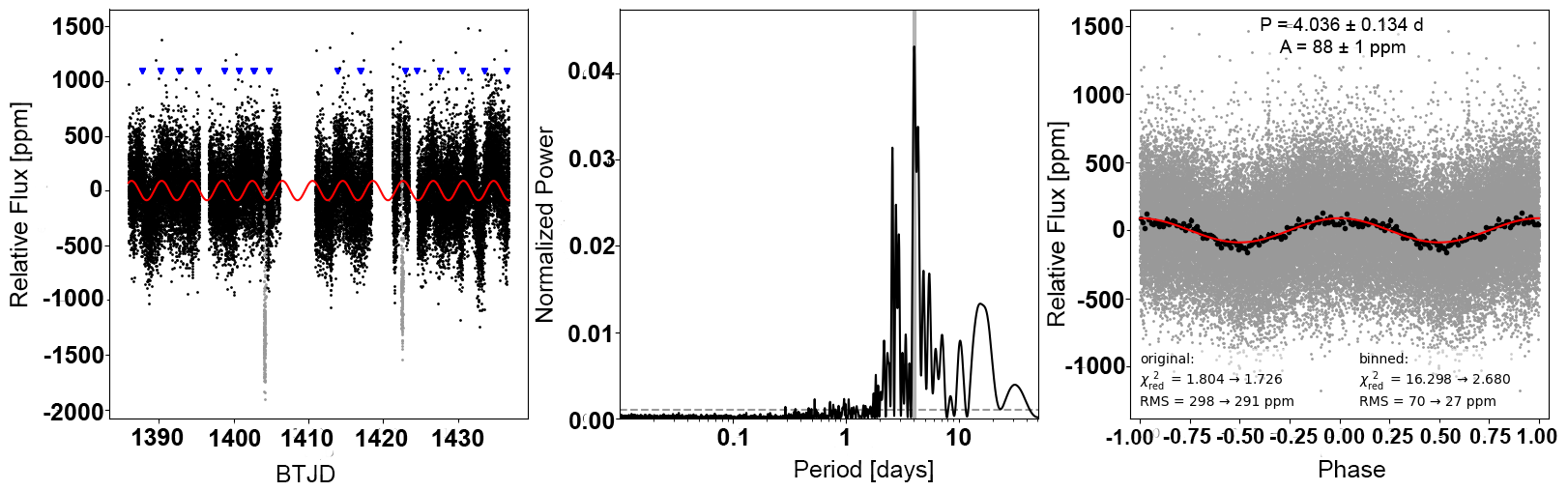}
  \caption{The left panel shows the \textit{TESS} light curve of TOI-257 from Sectors 3 and 4 with the best-fit variability. The middle panel is the Lomb-Scargle periodogram for raw light curves from Sectors 3 and 4 combined. The right panel is the phase-folded light curve at the peak period found from the Lomb-Scargle periodogram.}
  \label{lombscargle}
\end{figure*}

We calculated the Lomb-Scargle periodograms for the raw \textit{TESS} light curves from Sectors 3 and 4 individually and from the combined light curve of the two Sectors, after masking the transit events. For Sector 3, the periodogram shows that the variability has a period of $P=5.01\pm0.46$\,days and amplitude of $A=114\pm2$\,ppm. Sector 4 light curve has a variability with a period of $P=4.13\pm0.22$\,days and amplitude of $A=144\pm2$\,ppm. The period and amplitude from the two Sectors is reasonably consistent. Performing this analysis on the combined light curves reveals that the variability has a period of $P=4.04\pm0.13$\,d, amplitude of $A=88\pm1$\,ppm, and false alarm probability (FAP)$<<0.01$. A second very strong peak is observed at $\sim2.69$\,days (or $2P/3$) in the Lomb-Scargle periodograms. The FAP was computed from Monte Carlo simulations \citep[e.g.,][]{2010A&A...520A..15M} and the uncertainty in the period of variability was calculated following the procedure of \citet{2004A&A...417..557L}. The variability from both Sectors combined phases-up well at a period of 4.036\,days as shown in Figure~\ref{lombscargle}, which indicates that the variability is likely to be astrophysical in nature (from stellar rotation and star spots) and not systematics. We therefore have adopted the period of variability as $4.04\pm0.13$\,d.

We also performed an auto-correlation function analysis on the light curves from the individual Sectors and combined Sectors, and find that the period of variability as $P=5.03\pm0.61$\,days and $P=4.12\pm0.32$\,days for Sectors 3 and 4, respectively, and a period of $P=4.14\pm0.22$\,days for the combined light curves. We also find a strong secondary period in the combined light curves of $\sim2.7$\,days. These results are consistent with the periods found from the Lomb-Scargle periodograms.

To determine whether the period of variability is the true rotation period of the star or one of its harmonics, we calculate an upper limit on the rotation period from the star's $v\sin i$ and estimated radius. We measured the $v\sin i$ of TOI-257 by fitting a rotationally broadened Gaussian \citep{2005oasp.book.....G} to a least-squares deconvolution profile \citep{1997MNRAS.291....1D} obtained from the sum of all the spectral orders from the combined highest S/N {\sc {\textsc{Minerva}}}-Australis spectra of TOI-257. The resulting $v\sin i$ is $11.3\pm0.5$\,\kms\, and combined with the stellar radius from asteroseismology of $R_{\star}=1.888\pm0.033$\,\rsun, sets the upper limit on the rotation period for the star of $\sim8.5$\,days, assuming that the inclination of the stellar rotation axis is near $90\deg$ to the line of sight.

Given the above analyses, we attribute the $4.04$\,day period of variability observed in the combined \textit{TESS} light curve to be half the true rotation period of $8.08\pm0.26$\,days (which gives a $v_{rot}=2*\pi*R_{\star}/P_{rot}=11.8$\,\kms, consistent with the value of $v\sin i$). The very strong secondary peak observed at $\sim2.7$\,days in both the Lomb-Scargle periodograms and the auto-correlation function analysis provides further evidence in support of the $8.08\pm0.26$\,days being the true rotation period since the secondary peak corresponds nicely with the $P_{rot}/3$ harmonic. It is common for the observed rotational modulation to be at one or more of the harmonics, in particular at half and one-third the true rotation period \citep{vanderburg2016}. If the rotation period is $4.04$\,days, we would have expected to find a strong secondary peak at $\sim2.02$\,days instead of $\sim2.7$\,days. Given that the rotational period and stellar radius gives a rotational velocity consistent with the measured $v\sin i$, this suggest that the stellar obliquity is low (i.e., $i_{\star}\sim90\deg$).

%%%%%%%%%%%%%%%%%%%%%%%%%%%%%%%%%%%%%%%%%%%%%%%%%%%%%%%%%%%%%%%%%%%%

\subsection{Planetary System Parameters from Global Analysis}
\label{global_fit}
To determine the system parameters for TOI-257 and its planet, we used \texttt{EXOFASTv2} \citep{2013PASP..125...83E,2017ascl.soft10003E,2019arXiv190709480E} to perform a joint analysis of the \textit{TESS} photometry and the radial velocity data. We placed Gaussian priors on $T_{\rm eff}$ and [Fe/H] from the {\sc {\textsc{Minerva}}}-Australis high-resolution spectroscopy and Gaussian priors on $R_{\star}$ and $M_{\star}$ from asteroseismology. We applied an upper limit on the V-band extinction from the \citet{2011ApJ...737..103S} dust maps at the location of TOI-257. We also performed a separate SED analysis (so as not to double count information used from the asteroseismic priors) as an independent check on the stellar parameters using catalog photometry from Tycho \citep{2000AA...355L..27H}, 2MASS \citep{2003yCat.2246....0C}, WISE \citep{2013yCat.2328....0C}, and {\it Gaia\/} \citep{2018AA...616A...1G} as well as MIST stellar evolutionary models \citep{2016ApJS..222....8D,2016ApJ...823..102C}. Gaussian priors were placed on the parallax from {\it Gaia\/} DR2, adding 82\,$\mu$as to correct for the systematic offset found by \citet{2018ApJ...862...61S} and adding the 33\,$\mu$as uncertainty in their offset in quadrature to the {\it Gaia\/}-reported uncertainty. For the quadratic stellar limb darkening coefficients $u_{1}$ and $u_{2}$, \texttt{EXOFASTv2} applied weakly informative Gaussian priors drawn from the interpolation of the \citet{2011A&A...529A..75C} limb darkening models at each step in $\log{g}$, \teff, and [Fe/H] taken in the global model to help guide the coefficients. Table~\ref{stellar} lists the broadband magnitudes used in the SED analysis and the stellar parameters including the ones used as priors in the global analysis.

We ran two global models with \texttt{EXOFASTv2}, an eccentric orbit model with $e\cos{\omega_*}$ and $e\sin{\omega_*}$ as free parameters and a circular orbit model with eccentricity fixed to 0 to determine the significance of any potential eccentricity. We computed the small-sample Akaike Information Criterion (AICc) and Bayesian Information Criterion (BIC, see, \citealt{akaike1974, aicc2002}) for each model. We find that the $\Delta$\,AICc between the eccentric and circular model is 4.03, indicating that the circular model is moderately preferred over the eccentric model. Therefore, we have chosen the 1-planet circular model as the preferred solution. The results for analysis of both models are reported in Table~\ref{tab:TOI257.}.

\begin{figure}
  \includegraphics[width=8.5cm]{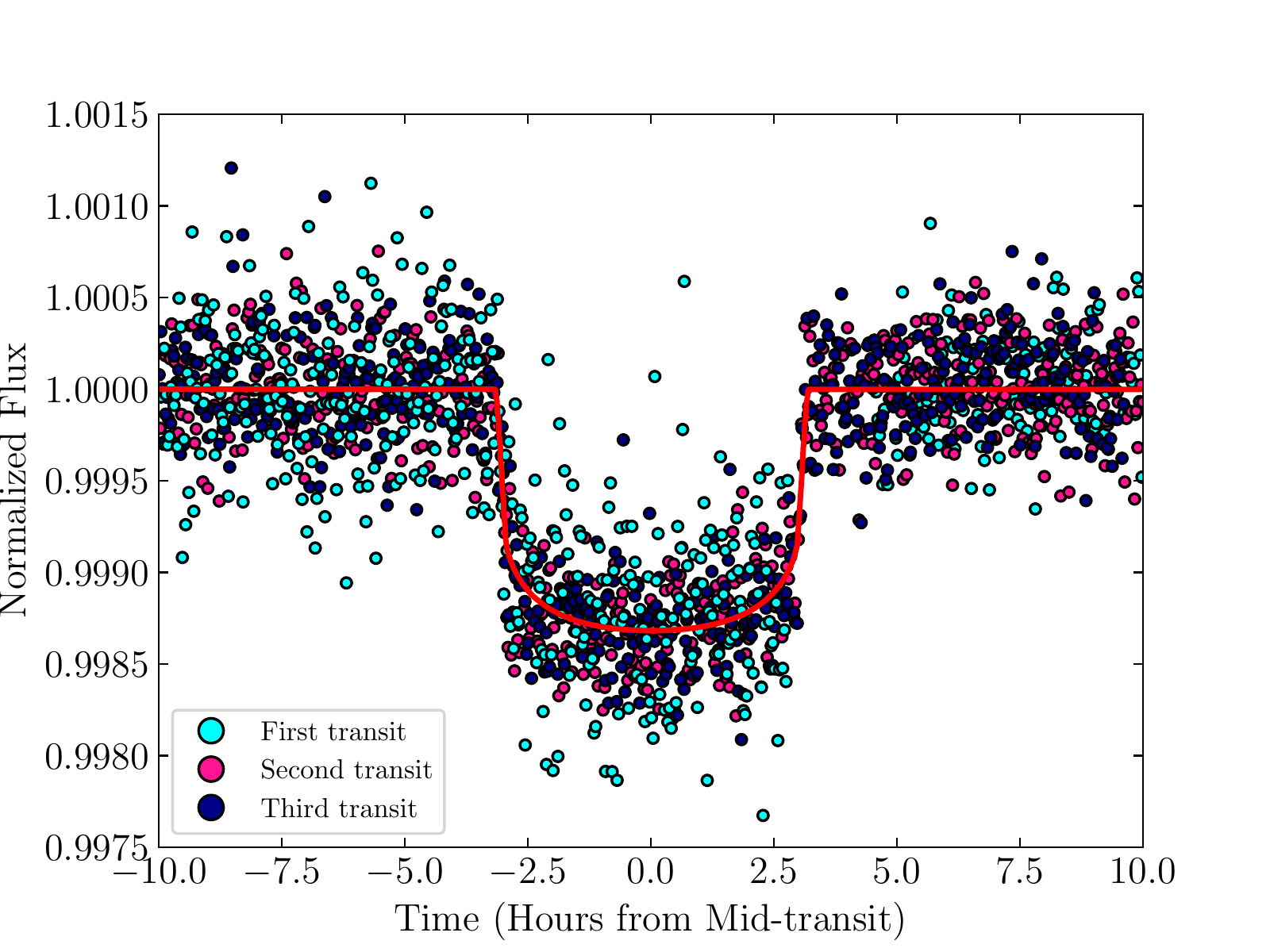}
  \caption{Phase folded \textit{TESS} light curve of TOI-257 with the individual transits color coded similar to Figure~\ref{tesstransits}. The red solid line is the best-fitting model.}
  \label{lc_phase_fig}
\end{figure}

\begin{figure}
  \includegraphics[width=8.5cm]{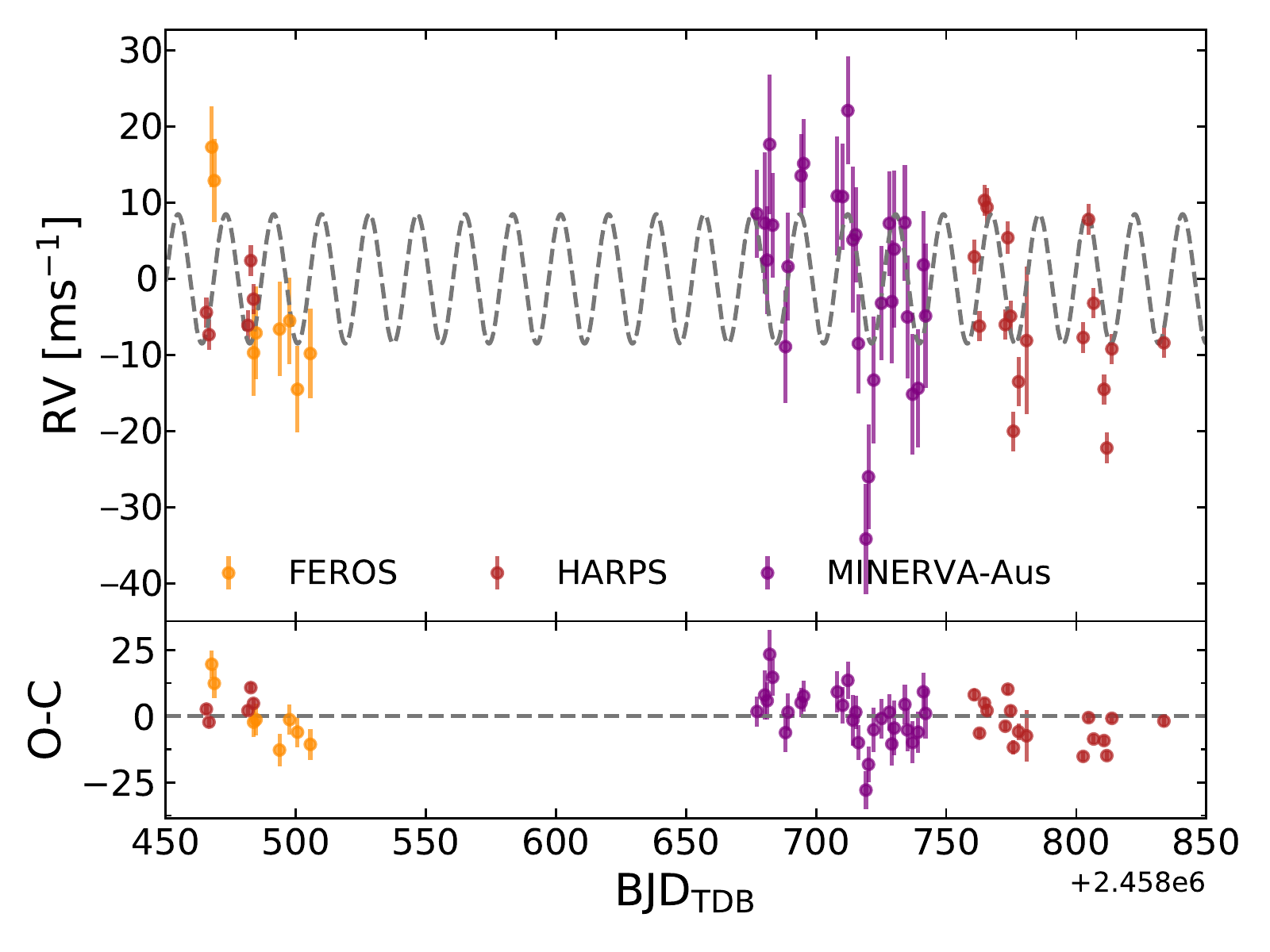}
  \caption{Radial velocity measurements of TOI-257 as a function of time. The radial velocity measurements from each instrument have been binned by day for clarity, however, the analysis was performed using the unbinned data. {\sc {\textsc{Minerva}}}-Australis radial velocities are represented by the purple filled-in circles. Radial velocities from FEROS and HARPS are the lime green and gold filled-in circles, respectively. The best-fit model is plotted as the dashed grey line and the center-of-mass velocity has been subtracted. The bottom panel shows the residuals between the data and the best-fit model.}
  \label{rv_full_fig}
\end{figure}

\begin{figure}
  \includegraphics[width=8.5cm]{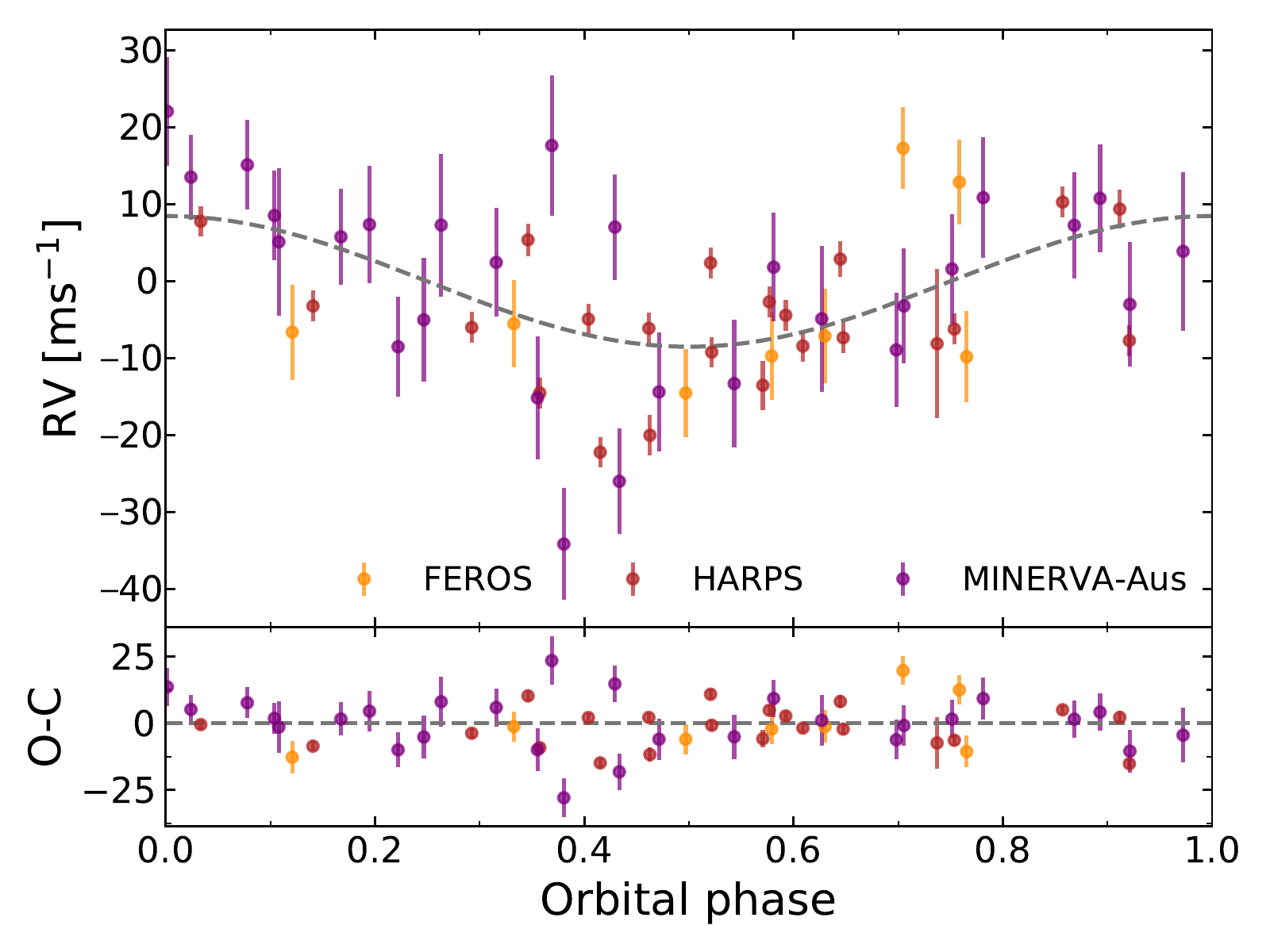}
  \caption{Same as Figure~\ref{rv_full_fig} but phased to one orbital period. The units of the horizontal axis were chosen so that the transit mid-time corresponds to an orbital phase of 0.25.}
  \label{rv_phase_fig}
\end{figure}

\begin{figure}
  \includegraphics[width=8.5cm]{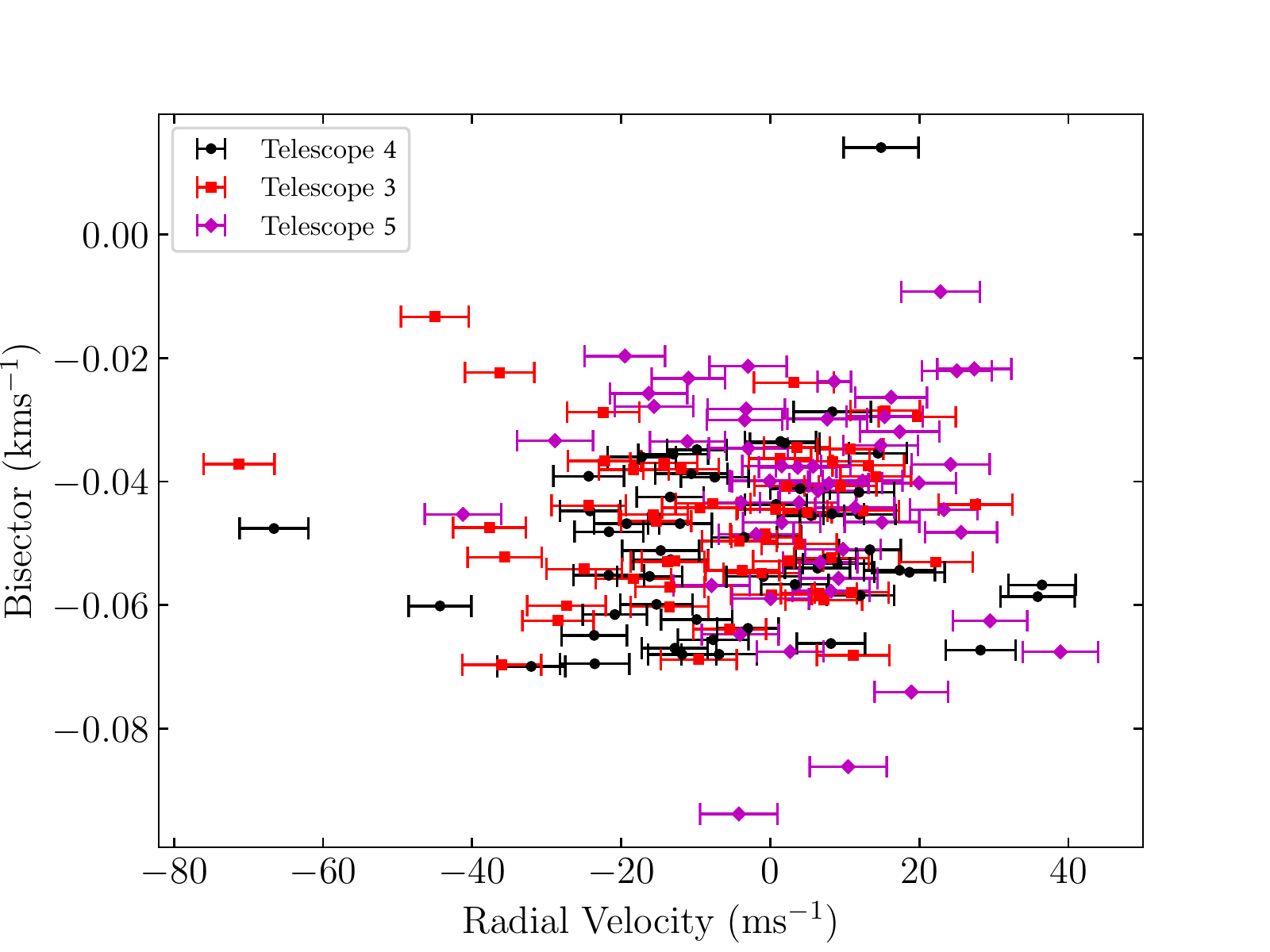}
  \caption{Bisector velocity span as a function of the radial velocities for the {\sc {\textsc{Minerva}}}-Australis radial velocities. There is no significant correlation between the bisector velocity span and the radial velocities.}
  \label{bvs_fig}
\end{figure}

The resulting best-fit models for the transit light curves are plotted in Figure~\ref{lc_phase_fig}, and for the radial velocities in Figures~\ref{rv_full_fig} and \ref{rv_phase_fig}. Figure~\ref{bvs_fig} is a plot of the bisector velocity span showing no correlation between the bisectors and the radial velocities for the {\textsc{Minerva}}-Australis observations, indicating that the measured radial velocity signal is likely planetary in nature and not due to stellar photospheric activity \citep{2013A&A...557A..93F}.

From the best-fit Kurucz stellar atmosphere model from the SED and the best-fitting MIST stellar evolutionary model, we find that TOI-257 is a somewhat evolved late-F star with $R_{\star}=1.951^{+0.066}_{-0.051}$\,\rsun, $M_{\star}=1.35^{+0.12}_{-0.38}$\,\msun, $T_{\rm eff}=6066^{+86}_{-110}$\,K, and $\log{g}=3.986^{+0.047}_{-0.150}$ (where $g$ is in units of \logg). These stellar parameters are in good agreement with the parameters derived from the {\textsc{Minerva}}-Australis and HARPS spectroscopy and asteroseismology. However, we choose not to adopt these stellar parameters since they are not as precise as the ones derived from spectroscopy and asteroseismology and list the stellar parameters derived from the joint analysis of the \textit{TESS} photometry and the radial velocity data in Table~\ref{tab:TOI257.}. From the joint analysis, we find that TOI-257 hosts a sub-Saturn sized planet with a radius of $R_P=0.639\pm0.013$\,$\rm{R_J}$\ ($7.16\pm0.15$\,\re) and mass of $M_P=0.138\pm0.023$\,$\rm{M_J}$\ ($43.9\pm7.3$\,\me), on a circular $\sim18.4$\, day orbit.

To ensure that our results are not potentially biased from the use of the \citet{2011A&A...529A..75C} limb darkening interpolation tables in \texttt{EXOFASTv2}, we ran two additional transit only circular models, one with and one without the Claret and Bloemen tables (setting the \texttt{NOCLARET} flag in \texttt{EXOFASTv2} such that $u_{1}$ and $u_{2}$ are completely free parameters). Both transit only models provided consistent results confirming the stellar and planetary parameters are not being biased by the \texttt{EXOFASTv2} Claret and Bloemen interpolation tables.

Whilst we do find that the circular model provides a somewhat better fit to the \textit{TESS} light curve and radial velocity data than compared with the eccentric model, the planet could still be on an eccentric orbit given the $\sim3.7$\,$\sigma$ eccentricity detection. As such, future transit observations to measure chromatic limb darkening as well as additional radial velocities can validate (or refute) any potential eccentricity in the orbit of TOI-257\,b.

\begin{table*}
\caption{Median values and 68\% confidence interval for TOI-257 from the MCMC \texttt{EXOFASTv2} analysis of both eccentric and circular orbital models. \textbf{Notes.} -- M-A T3, M-A T4, and M-A T6 are {\textsc{Minerva}}-Australis Telescope3, Telescope4, and Telescope6, respectively.
$^{\ast}$The time of conjunction that is closest to the starting value supplied as a prior and is typically a good approximation for the mid transit time.
%$^{\diamond}$The optimal conjunction time is the time of conjunction that minimizes the covariance with the orbital period and therefore has the smallest uncertainty.
$^{\wedge}$The equilibrium temperature of the planet assumes no albedo and perfect heat redistribution.
$^{\star}$The tidal circularization timescale is calculated using Equation 3 from \citet{2006ApJ...649.1004A} and assuming a $Q=10^{6}$.
$^{\dagger}$\textit{TESS} LC1 is the \textit{TESS} light curve from PDC and \textit{TESS} LC2 is the \textit{TESS} light curve produced using the \citet{2019ApJ...881L..19V} procedures.}
\begin{center}
\begin{minipage}{\linewidth}
\begin{tabularx}{\textwidth}{lcccccc}
\hline\hline
~~~Parameter & Description & Eccentric Model & Circular Model \\
\hline \\
%\vspace{0.5pt} \\
%\smallskip \\
\multicolumn{3}{l}{Stellar Parameters:} & & \smallskip \\
~~~~$M_*$\dotfill &Mass (\msun)\dotfill &$1.394\pm0.046$ & $1.407^{+0.045}_{-0.046}$\\
~~~~$R_*$\dotfill &Radius (\rsun)\dotfill &$1.883\pm0.033$ & $1.867^{+0.033}_{-0.032}$\\
~~~~$L_*$\dotfill &Luminosity (\lsun)\dotfill &$4.41^{+0.31}_{-0.29}$ & $4.33^{+0.30}_{-0.28}$\\
~~~~$\rho_*$\dotfill &Density (cgs)\dotfill &$0.294^{+0.019}_{-0.017}$ &$0.305^{+0.019}_{-0.018}$\\
~~~~$\log{g}$\dotfill &Surface gravity (cgs)\dotfill &$4.032\pm0.021$ &$4.044\pm0.021$\\
~~~~$T_{\rm eff}$\dotfill &Effective Temperature (K)\dotfill &$6096\pm89$ &$6095\pm89$\\
~~~~$[{\rm Fe/H}]$\dotfill &Metallicity (dex)\dotfill &$0.177\pm0.099$ &$0.175^{+0.099}_{-0.098}$\\
%\vspace{0.5pt} \\
\multicolumn{2}{l}{Planetary Parameters:}  & \multicolumn{2}{c}{b} \smallskip \\
~~~~$P$\dotfill &Period (days)\dotfill &$18.38827\pm0.00072$ &$18.38818^{+0.00085}_{-0.00084}$\\
~~~~$R_P$\dotfill &Radius (\rj)\dotfill &$0.626^{+0.013}_{-0.012}$ &$0.639\pm0.013$\\
~~~~$M_P$\dotfill &Mass (\mj)\dotfill &$0.134^{+0.023}_{-0.022}$ &$0.138\pm0.023$\\
~~~~$T_C$\dotfill &Time of conjunction (\bjdtdb)\dotfill &$2458385.7600\pm0.0011$ $^{\ast}$ &$2458385.7601^{+0.0013}_{-0.0012}$ $^{\ast}$\\
%~~~~$T_0$\dotfill &Optimal conjunction Time (\bjdtdb)\dotfill &$2458404.14831\pm0.00056$$^{\diamond}$\\
~~~~$a$\dotfill &Semi-major axis (AU)\dotfill &$0.1523\pm0.0017$ &$0.1528^{+0.0016}_{-0.0017}$\\
~~~~$i$\dotfill &Inclination (Degrees)\dotfill &$88.78^{+0.78}_{-0.57}$ &$87.91^{+0.11}_{-0.10}$\\
~~~~$e$\dotfill &Eccentricity \dotfill &$0.242^{+0.040}_{-0.065}$ &0 (fixed)\\
~~~~$\omega_*$\dotfill &Argument of Periastron (Degrees)\dotfill &$96\pm22$ & ... \\
~~~~$T_{eq}$\dotfill &Equilibrium temperature (K)\dotfill &$1033^{+19}_{-18}$ $^{\wedge}$ &$1027\pm18$\\
~~~~$\tau_{\rm circ}$\dotfill &Tidal circularization timescale (Gyr)\dotfill &$1880^{+400}_{-360}$ $^{\star}$ &$1750^{+380}_{-330}$ $^{\star}$\\
~~~~$K$\dotfill &RV semi-amplitude (\mos)\dotfill &$8.5\pm1.4$ &$8.5\pm1.4$\\
~~~~$R_P/R_*$\dotfill &Radius of planet in stellar radii \dotfill &$0.03414^{+0.00037}_{-0.00029}$ &$0.03521\pm0.00022$\\
~~~~$a/R_*$\dotfill &Semi-major axis in stellar radii \dotfill &$17.39^{+0.36}_{-0.35}$ &$17.60^{+0.36}_{-0.35}$\\
~~~~$\delta$\dotfill &Transit depth (fraction)\dotfill &$0.001166^{+0.000025}_{-0.000020}$ &$0.001240\pm0.000016$\\
~~~~$\tau$\dotfill &Ingress/egress transit duration (days)\dotfill &$0.00947^{+0.00160}_{-0.00072}$ &$0.01532^{+0.00070}_{-0.00068}$\\
~~~~$T_{14}$\dotfill &Total transit duration (days)\dotfill &$0.2644^{+0.0017}_{-0.0013}$ &$0.2702\pm0.0013$\\
%~~~~$T_{FWHM}$\dotfill &FWHM transit duration (days)\dotfill &$0.2548\pm0.0010$\\
~~~~$b$\dotfill &Transit Impact parameter \dotfill &$0.28^{+0.17}_{-0.19}$ &$0.643^{+0.019}_{-0.020}$\\
~~~~$\rho_P$\dotfill &Density (cgs)\dotfill &$0.67^{+0.13}_{-0.12}$ &$0.65^{+0.12}_{-0.11}$\\
~~~~$logg_P$\dotfill &Surface gravity (cgs) \dotfill &$2.927^{+0.072}_{-0.081}$ &$2.922^{+0.070}_{-0.079}$\\
~~~~$\Theta$\dotfill &Safronov Number \dotfill &$0.0467^{+0.0080}_{-0.0078}$ &$0.0469^{+0.0078}_{-0.0076}$\\
~~~~$\fave$\dotfill &Incident Flux (\fluxcgs)\dotfill &$0.245^{+0.018}_{-0.017}$ &$0.253^{+0.019}_{-0.017}$\\
~~~~$T_P$\dotfill &Time of Periastron (\bjdtdb)\dotfill &$2458367.59^{+0.68}_{-0.69}$ &$2458385.7601^{+0.0013}_{-0.0012}$\\
%~~~~$T_A$\dotfill &Time of Ascending Node (\bjdtdb)\dotfill &$2458382.34^{+0.43}_{-0.54}$\\
%~~~~$T_D$\dotfill &Time of Descending Node (\bjdtdb)\dotfill &$2458370.58^{+0.48}_{-0.40}$\\
~~~~$e\cos{\omega_*}$\dotfill & \dotfill &$-0.026^{+0.083}_{-0.084}$ & ... \\
~~~~$e\sin{\omega_*}$\dotfill & \dotfill &$0.225^{+0.040}_{-0.068}$ & ... \\
~~~~$M_P/M_*$\dotfill &Mass ratio \dotfill &$0.000092^{+0.000016}_{-0.000015}$ &$0.000094^{+0.000016}_{-0.000015}$\\
~~~~$d/R_*$\dotfill &Separation at mid transit \dotfill &$13.42^{+1.20}_{-0.88}$ &$17.60^{+0.36}_{-0.35}$\\
%\vspace{0.5pt} \\
\multicolumn{2}{l}{Wavelength Parameters:} & \textit{TESS} (Eccentric Model) & \textit{TESS} (Circular Model) \smallskip \\
~~~~$u_{1}$\dotfill &linear limb-darkening coeff \dotfill &$0.222\pm0.031$ &$0.221\pm0.032$\\
~~~~$u_{2}$\dotfill &quadratic limb-darkening coeff \dotfill &$0.274\pm0.034$ &$0.268\pm0.033$\\
~~~~$A_D$\dotfill &Dilution from neighboring stars \dotfill &$\leq0.00053$ &$\leq0.00052$\\
%\vspace{0.5pt} \\
\multicolumn{2}{l}{Telescope Parameters (Eccentric Model):} & FEROS & HARPS & M-A T3 & M-A T4 & M-A T6 \smallskip \\
~~~~$\gamma_{\rm rel}$\dotfill &Relative RV Offset (\mos)\dotfill &$-5.6\pm5.3$&$28.7\pm1.5$&$1.0\pm2.0$&$0.1\pm2.0$&$-0.8\pm2.1$\\
~~~~$\sigma_J$\dotfill &RV Jitter (\mos)\dotfill &$13.3^{+6.9}_{-4.3}$&$7.07^{+1.20}_{-0.94}$&$13.9^{+1.7}_{-1.4}$&$14.0^{+1.7}_{-1.4}$&$13.9^{+1.8}_{-1.5}$\\
~~~~$\sigma_J^2$\dotfill &RV Jitter Variance \dotfill &$175^{+230}_{-95}$&$49^{+18}_{-12}$&$194^{+49}_{-37}$&$197^{+50}_{-38}$&$193^{+53}_{-40}$\\
%\vspace{0.5pt} \\
\multicolumn{2}{l}{Telescope Parameters (Circular Model):} & FEROS & HARPS & M-A T3 & M-A T4 & M-A T6 \smallskip \\
~~~~$\gamma_{\rm rel}$\dotfill &Relative RV Offset (\mos)\dotfill &$-4.8\pm5.3$&$29.3\pm1.5$&$0.8\pm2.0$&$-0.1\pm2.0$&$-1.0\pm2.1$\\
~~~~$\sigma_J$\dotfill &RV Jitter (\mos)\dotfill &$13.1^{+6.9}_{-4.2}$&$7.29^{+1.20}_{-0.98}$&$14.0^{+1.7}_{-1.4}$&$13.8^{+1.7}_{-1.4}$&$13.8^{+1.8}_{-1.5}$\\
~~~~$\sigma_J^2$\dotfill &RV Jitter Variance \dotfill &$170^{+230}_{-92}$&$53^{+19}_{-13}$&$195^{+50}_{-38}$&$190^{+49}_{-37}$&$189^{+52}_{-40}$\\
%\vspace{0.5pt} \\
\multicolumn{2}{l}{Transit Parameters:}& \textit{TESS} LC1$^{\dagger}$ & \textit{TESS} LC2$^{\dagger}$ \smallskip \\
~~~~$\sigma^{2}$\dotfill &Added Variance \dotfill &$1.70\pm0.12\times10^{-8}$&$1.33^{+0.60}_{-0.56}\times10^{-8}$\\
~~~~$F_0$\dotfill &Baseline flux \dotfill &$1.0000009\pm0.0000034$&$1.000001\pm0.000010$\\
%\vspace{0.5pt} \\
\multicolumn{2}{l}{Model Comparison Statistics:}& Eccentric Model & Circular Model \smallskip \\
~~~$\Delta$\,AICc\dotfill &Akaike Information Criterion \dotfill &4.03&0\\
~~~~$\Delta$\,BIC\dotfill &Bayesian Information Criterion \dotfill &17.34&0\\
\hline
\end{tabularx}
\end{minipage}
\end{center}
\label{tab:TOI257.}
\end{table*}

\subsection{Complementary Analysis, and Limits on Additional Planets}
\label{limits}
We further analyze the radial velocity data set in Table~\ref{vels} with \texttt{RadVel} \citep{fulton2018a} to provide both an independent analysis for checking consistency in the mass and eccentricity of planet b, and to search for any additional planets.  The search for additional planets is motivated by two reasons.  First, moderately eccentric Keplerian signals can sometimes resolve into two near-circular resonant signals with additional radial velocity data \citep[e.g.][]{witt13,trifonov17,boisvert18,witt19}. Second, we wish to evaluate the multiplicity of systems like TOI-257 with warm sub-Saturns. 

The combined analysis of the HARPS, \textsc{Minerva}-Australis, and FEROS data sets are consistent with a planet at the known transiting period and $T_{c}$ from Table~\ref{tab:TOI257.}. The circular orbital solution is marginally favored over an eccentric model (in agreement with the \texttt{ExoFASTv2} analysis), according to the relative small-sample Akaike Information Criterion ($\Delta \textrm{AICc}=3.14$). We fix $P$ and $T_{c}$ to the values independently derived from the \texttt{ExoFASTv2} analysis of the \textit{TESS} light curve as they will not be well-constrained from the RVs alone, considering the small baseline compared to the orbital period. The best fit semi-amplitude from \texttt{RadVel} is similar to the \texttt{EXOFASTv2} result, $8.6\pm1.2$\,\mos.

%Since we do not jointly model the light curve with the radial velocities in \texttt{RadVel}, we are not capturing the additional information from the transit duration ($\sim7.804$\,hr for circular orbit versus the measured $6.346$\,hr duration) and shape present in the light curve that may contribute to the increased evidence for an eccentric orbital solution presented earlier with \texttt{ExoFASTv2}. The best fit semi-amplitude from \texttt{RadVel} is similar to the \texttt{EXOFASTv2} result, $8.6\pm1.2$\,\mos.

\begin{figure}
  \includegraphics[width=8.5cm]{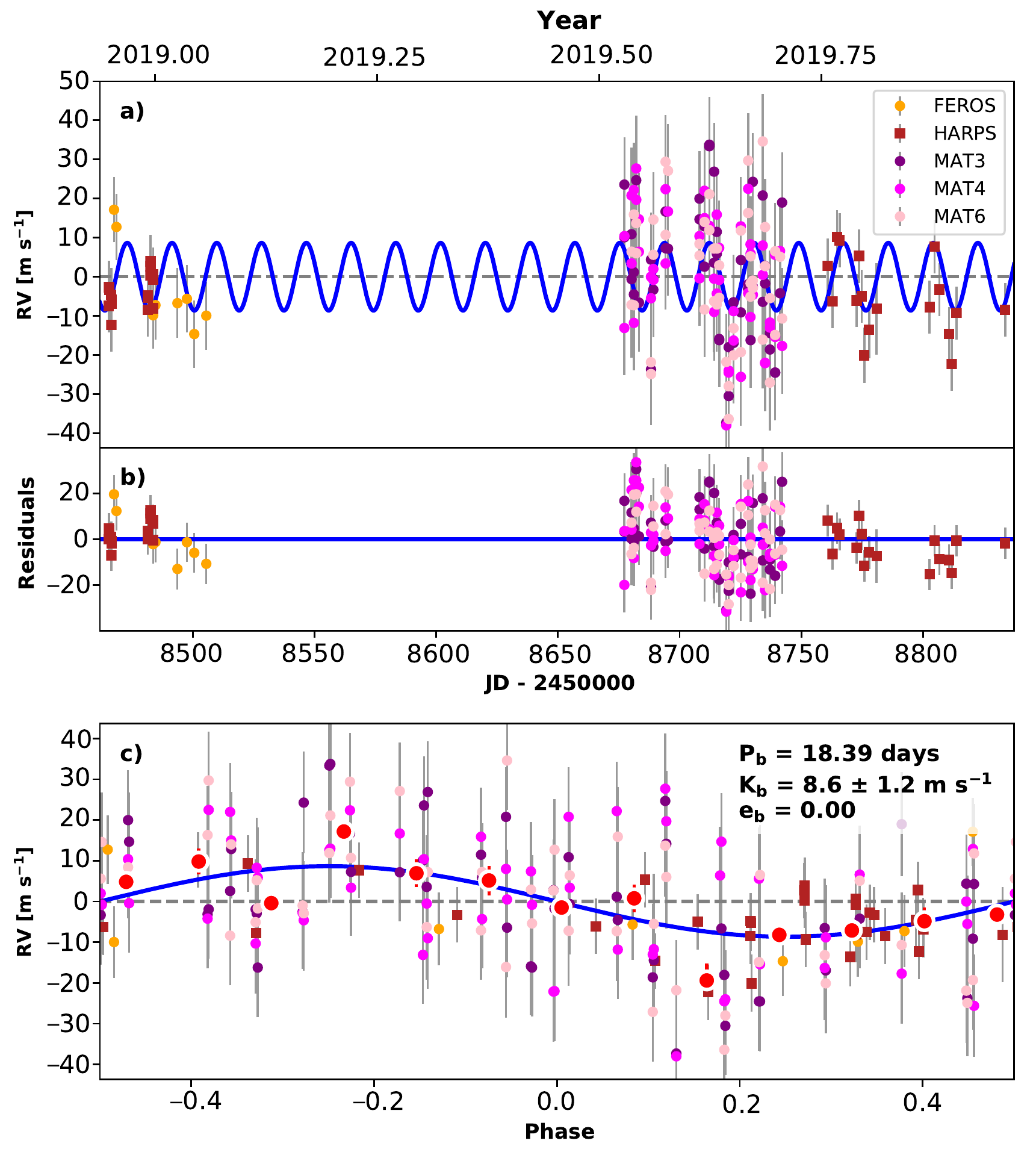}
  \caption{\textbf{a)} Best-fit 1-planet Keplerian orbital model for TOI-257. The maximum likelihood model is plotted in blue. We add in quadrature the radial velocity jitter terms listed in Table~\ref{tab:TOI257.} with the measurement uncertainties for all radial velocities to determine individual error bars. \textbf{b)} Residuals to the best fit 1-planet model. \textbf{c)} Radial velocities phase-folded to the period of planet b. Red circles are the individual velocities binned in 0.08 units of orbital phase.}
  \label{fig:radvel_1}
\end{figure}

The remaining scatter in the residuals after removing planet b from the one-planet orbital solution is consistently larger than the measured uncertainties of the three instruments and appears structured (see Figure~\ref{fig:radvel_1}). We use a custom modified version of \texttt{RadVel} to generate log-likelihood periodograms (LLPs) with various orbit assumptions to search for additional planets. We start with a single planet model and generate a log-likelihood for a wide range in fixed periods, fitting only for $T_{c}$ and $K$, as well as the relative instrument dependent offsets and additional radial velocity ``jitter'' noise terms, and then a second LLP assuming a fixed period and $T_{c}$ for planet b, but varying both semi-amplitudes to search for an additional planet candidate TOI-257c. Anecdotally, we observe that allowing for eccentric orbits in LLPs typically results in a noisier LLP compared to considering only circular orbits, and can particularly yield false peaks where $e\approx1$ with the region of largest |$\mathrm{d}RV$/$\mathrm{d}t$| located where the radial velocities are minimally sampled. These are likely non-physical orbits, so we only present circular searches (similarly, considering only eccentricities $\lesssim0.5$ mitigates this effect). 

\begin{figure}
  \includegraphics[width=8.5cm]{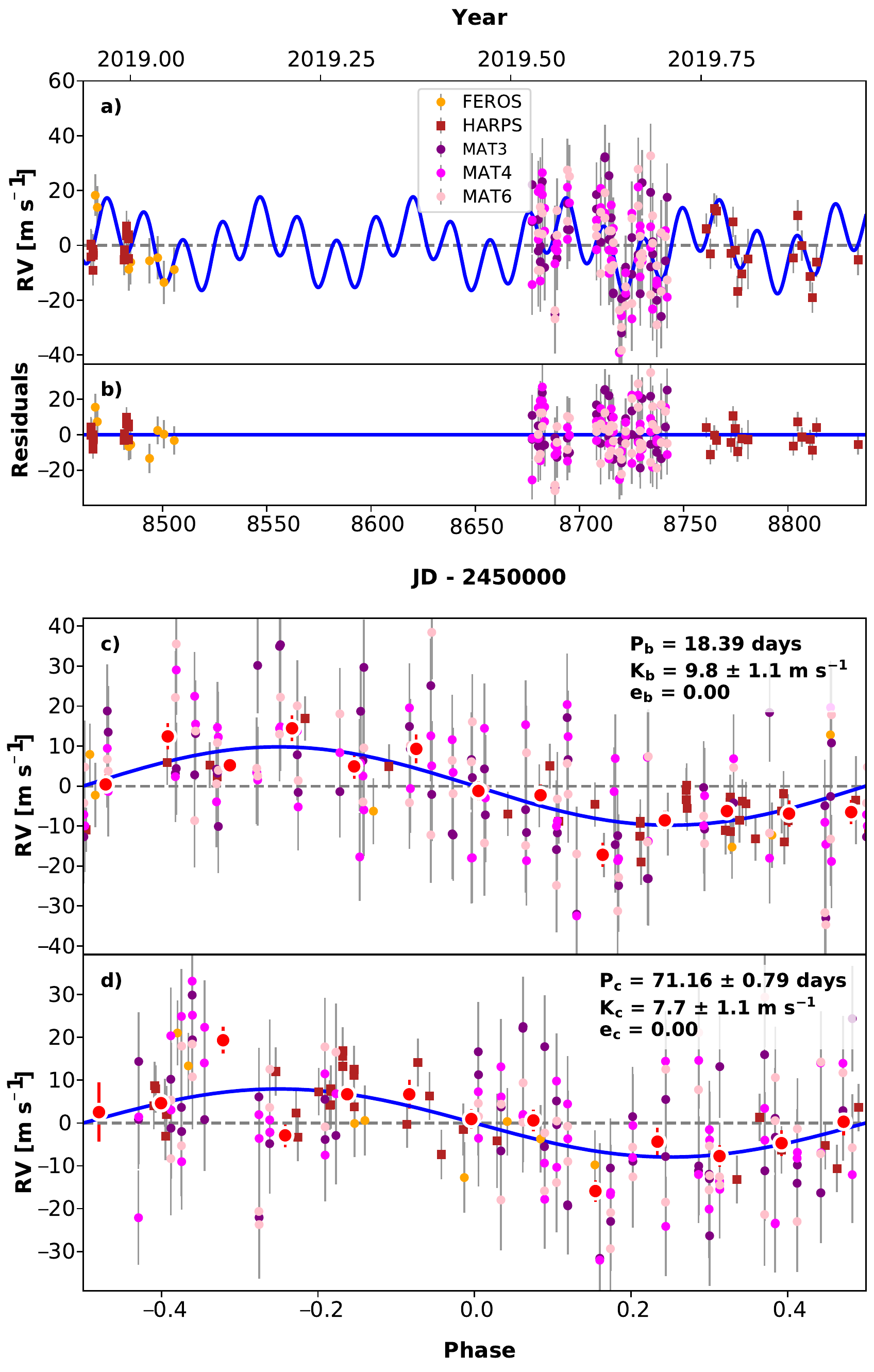}
  \caption{\textbf{a)} Best-fit 2-planet Keplerian orbital model for TOI-257. The maximum likelihood model(s) is plotted in blue. We add in quadrature the radial velocity jitter terms listed in Table~\ref{tab:TOI257.} with the measurement uncertainties for all radial velocities to determine individual error bars. \textbf{b)} Residuals to the best fit 2-planet model. \textbf{c)} Same, but radial velocities phase-folded to the period of planet b. \textbf{d)} Same, but radial velocities phase-folded to the period of a possible planet c. Red circles (if present) are the individual velocities binned in 0.08 units of orbital phase.}
  \label{fig:radvel_2}
\end{figure}

Both the single planet model LLP and the two planet model LLP feature a peak near 71 days (Figure~\ref{fig:radvel_3_freq}). Including a circular planet near the 71 day peak, with no prior on $T_{c}$, yields a posterior probability distribution of the semi-amplitude for the second planet that is $7\sigma$ deviant from 0, and minimally affects the statistical significance of the first planet (as shown in Figure~\ref{fig:radvel_2}). The model comparison heavily favors the 2-planet model over the 1-planet model with $\Delta \mathrm{AICc}=38.10$ (evidence ratio of $1.88\times10^{8}$). This 71 day signal translates to approximately a $0.2\%$ transit depth assuming the mass-radius relation given by \cite{chen2017} and stellar parameters in Table~\ref{tab:TOI257.}. Posterior distributions plots from \texttt{RadVel} for a 1-planet and 2-planet circular models are available as supplementary material online.

%Figures~\ref{fig:radvel_4} and \ref{fig:radvel_5} in the Appendix shows the posterior distributions from \texttt{RadVel} for a 1-planet and 2-planet circular models, respectively. 

As a consistency check with the 2-planet preferred solution with \texttt{RadVel}, we ran an additional global model using \texttt{EXOFASTv2} that included fitting both planet b and c, with eccentricity fixed to 0 for both planets and using the same priors as before in our 1-planet analysis. We used uniform priors on $T_{c}$ and $P$ for planet c, with the starting values on those parameters from the best fit values found with \texttt{RadVel}. We see no evidence for a transit in the \textit{TESS} light curve within the uncertainty window of the best fit $T_{c}$ for the possible outer planet with \texttt{RadVel} and have therefore excluded fitting transits for planet c in this analysis. The analysis with \texttt{EXOFASTv2} is unable to constrain the orbital parameters for planet c, resulting in a best-fit period of $P=378^{+98}_{-310}$\,days and time of conjunction of $T_{c}=2458709.7^{+20.0}_{-3.9}$\,BJD. The best-fit $K$ is $9.8^{+7.6}_{-2.8}$\,\mos. The orbital and planetary parameters for planet b remain consistent with the best-fit values of the 1-planet circular orbital model reported in Table~\ref{tab:TOI257.}. A comparison of the AICc and BIC between the 2-planet and 1-planet models gives a $\Delta$\,AICc of 11.46 and $\Delta$\,BIC of 46.23, strongly favouring the 1-planet circular model. However, we note that the \texttt{EXOFASTv2} analysis was not able to reach full convergence in a reasonable amount of time and could be the source of discrepancy between \texttt{RadVel} and \texttt{EXOFASTv2}.

While this radial velocity detection is significant from the analysis with \texttt{RadVel}, more high-precision radial velocity measurements are needed to ensure the candidate c planet signal is not an alias or possibly a result of the observing cadence, especially without an observed transit event and the 2-planet model being disfavored in the \texttt{EXOFASTv2} analysis.

\subsection{Assessing the Level of Stellar Activity Present in the Radial Velocities}

\begin{figure}
  \includegraphics[width=8.5cm]{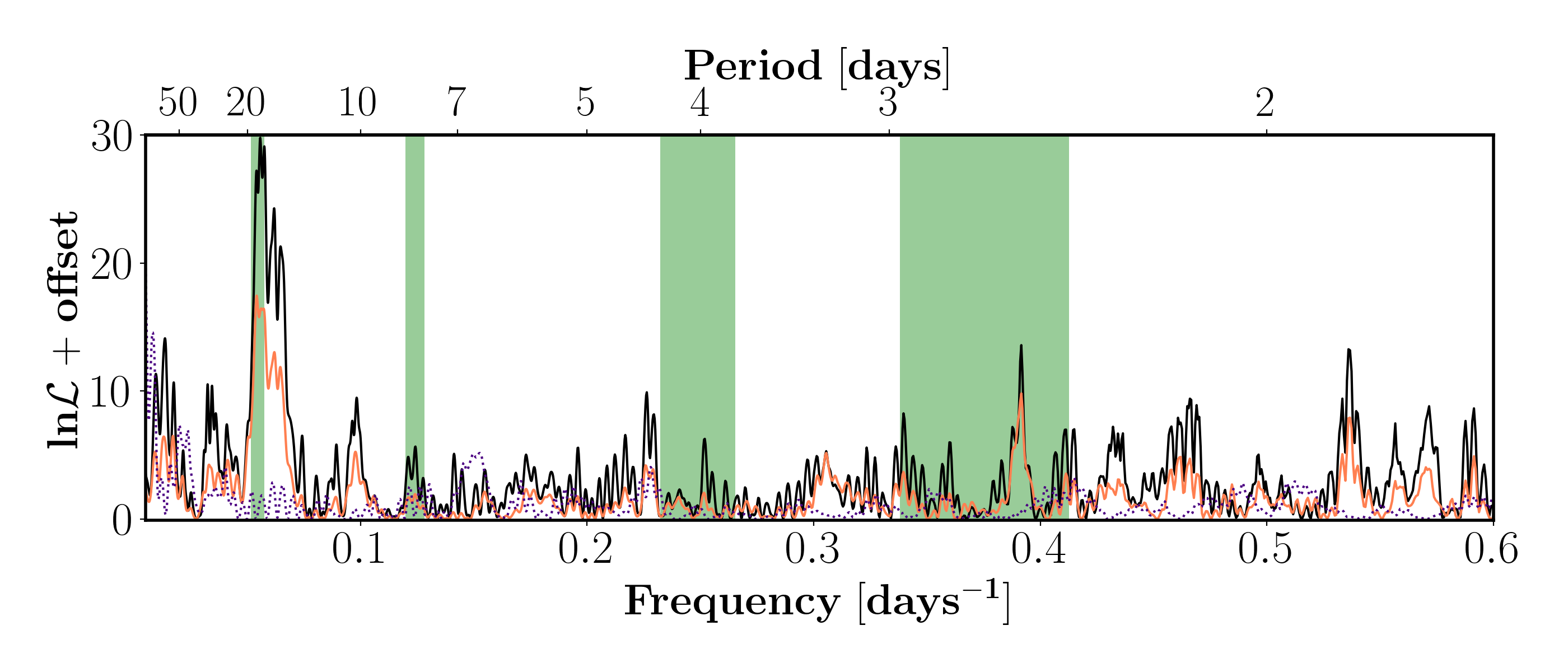}
  \includegraphics[width=8.5cm]{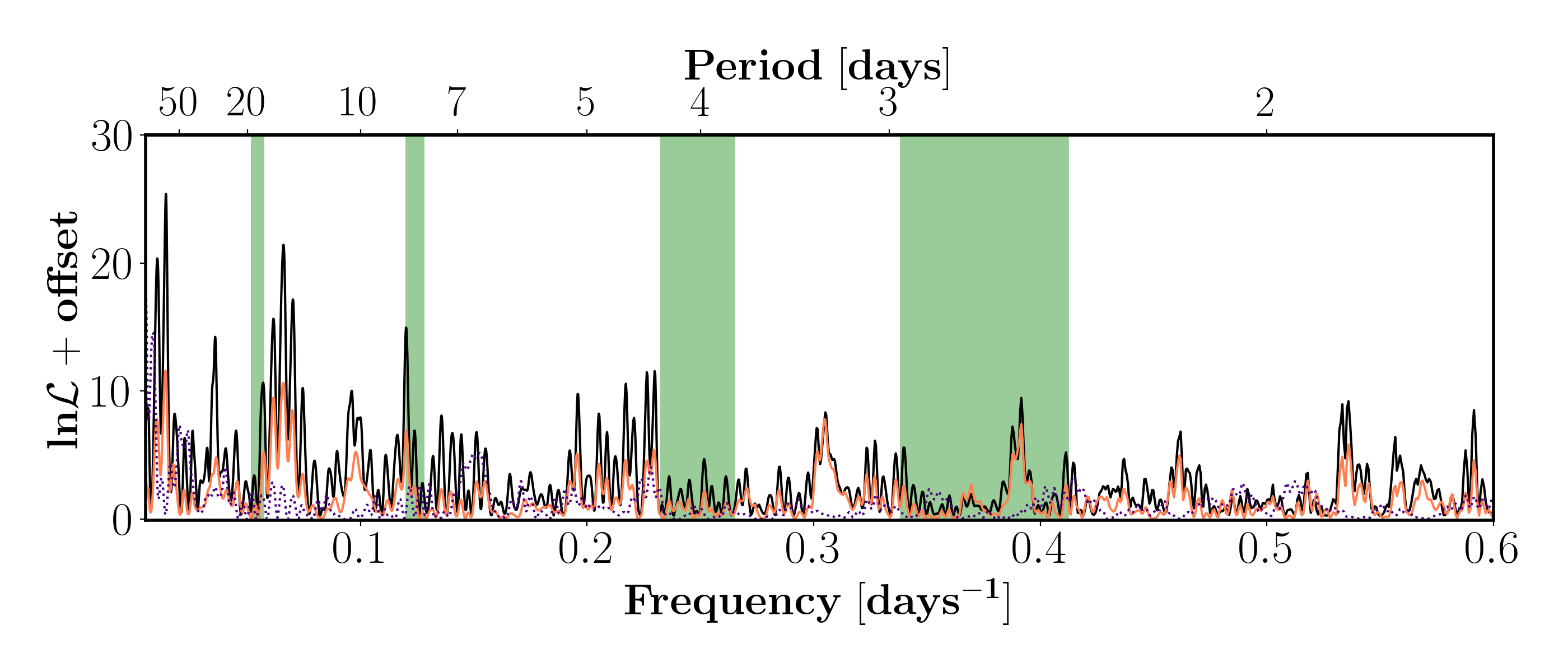}
  \includegraphics[width=8.5cm]{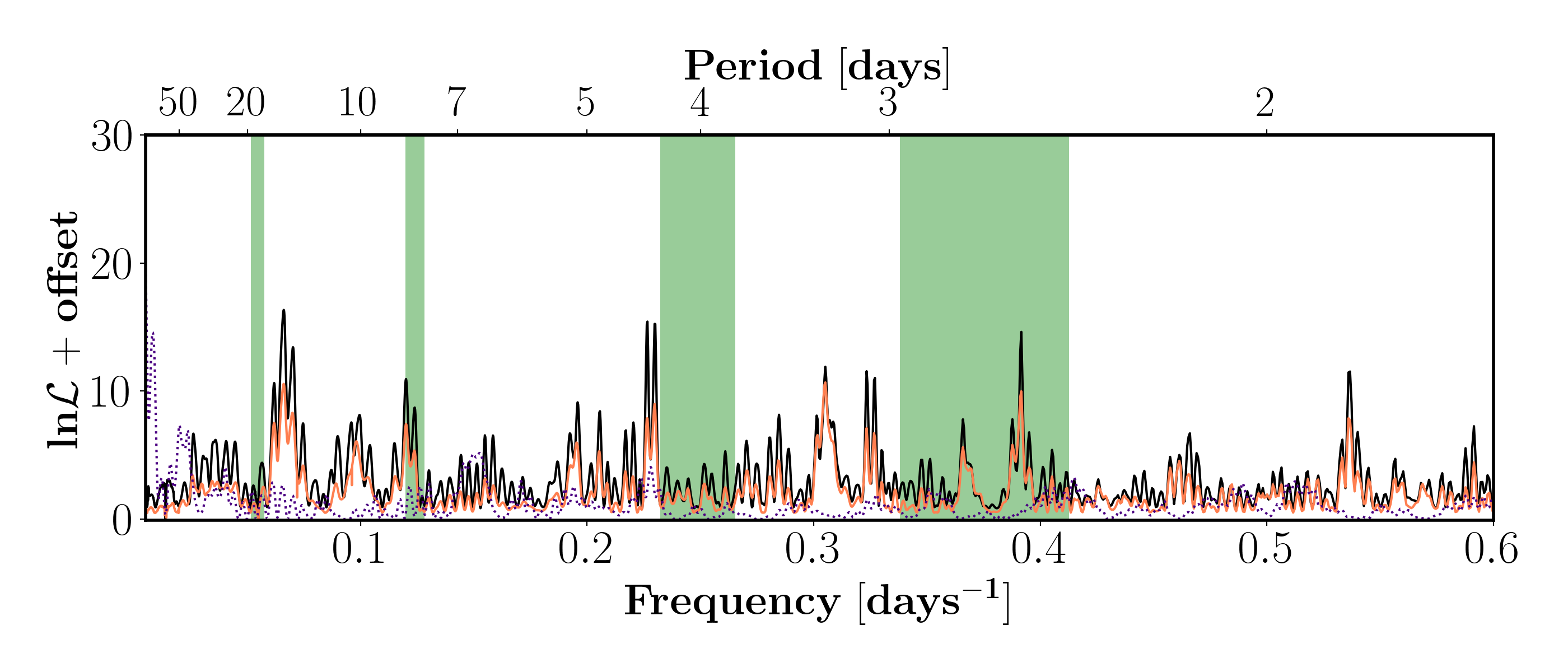}
  \caption{Log-likelihood periodograms as a function of frequency generated using \texttt{RadVel}. Shaded from left to right are the frequency of planet b (insignificant width), the estimated stellar rotation frequency $1/P_{\mathrm{rot}}$ from Table~\ref{hyperparams} (width is $\pm1\sigma$), and harmonics $2/P_{\mathrm{rot}}, 3/P_{\mathrm{rot}}$. The orange line includes a Gaussian Process (GP) to model stellar-activity with a quasi-periodic kernel with priors listed in Table~\ref{hyperparams}, while the black line models remaining jitter as per-instrument Gaussian noise. The dashed line represents the window function (arbitrary scaling). \textbf{Top}: A 1-planet circular model tested at a wide range in fixed periods, fitting for $K, T_{c}$ and the relative instrument dependent offsets and noise terms (or single GP). \textbf{Middle}: Same, but for a 2-planet model assuming a fixed period and $T_{c}$ for planet b from Table~\ref{tab:TOI257.}, but varying both semi-amplitudes to search for additional planets. \textbf{Bottom}: Same as middle, but searching for a third planet such that $\mathrm{P_{c}}\sim\mathcal{N}(71.5, 1)$. We see no evidence for additional periodic variations in the RVs past 18 days over the full observing window.}
  \label{fig:radvel_3_freq}
\end{figure}

Next, we consider the possibility that the excess radial velocity residuals after modeling planet b are due to stellar activity rather than a second planet (or both) as presented in the previous subsection. At this time, \texttt{EXOFASTv2} does not permit the inclusion of a stellar activity model for the radial velocities, whereas \texttt{RadVel} does. With our customized version of \texttt{RadVel}, we calculate LLPs using a Gaussian Process (GP) with a quasi-periodic kernel \citep{2015MNRAS.452.2269R}\footnote{The specific implementation of the quasi-periodic kernel in \texttt{RadVel} can be found on \url{https://radvel.readthedocs.io/en/latest/tutorials/GaussianProcess-tutorial.html}} to approximate any detectable stellar-activity. We re-run the MCMC analysis for 1- and 2-planet models. We assume broad Gaussian priors on the GP hyper-parameters listed in Table~\ref{hyperparams}. Both $\sim$4 or $\sim$8 day GP period produce qualitatively similar LLPs and mitigate peaks less than the candidate $P_{\mathrm{rot}}$ and show strong evidence for both the transiting planet and the candidate planet near 71~days (Figure~\ref{fig:radvel_3_freq}). The GP model is strongly favored in the 1-planet case, but only marginally so for the 2-planet models ($\Delta \mathrm{AICc}=18.68, 2.73$, respectively). However, a 2-planet model with a GP is still favored over the corresponding 1-planet model ($\Delta \textrm{AICc}=21.76$). A summary of the information criteria for the tested models is provided in table \ref{tab:info_criteria}. Posterior distribution plots from \texttt{RadVel} with a quasi-periodic Gaussian Process for a 1-planet and 2-planet circular models are available online as supplementary material.

%Figures~\ref{fig:radvel_6} and \ref{fig:radvel_7} in the Appendix shows the posterior distributions from \texttt{RadVel} with a quasi-periodic Gaussian Process for a 1-planet and 2-planet circular models, respectively.

% old sentence
% ... favor a single per-instrument Gaussian noise model over a $\sim$ 4 day GP ($\Delta \textrm{AICc}=3.22$), while the GP is only marginally favored for the 8~day period case ($\Delta \textrm{AICc}=0.72$)

Despite being statistically favored ($\sim5.1\sigma$ detection) with \texttt{RadVel}, we do not claim TOI-257c as a confirmed planet. \cite{nava2019} has shown that activity can introduce spurious periodogram peaks at orbital periods longer than the stellar rotation period over the course of a single season, particularly for radial velocities that are unevenly sampled as is the case herein, notably for the HARPS data. However, with adequately sampled data (densely sampled with nightly cadence), \cite{vanderburg2016} find no evidence of spurious radial velocity periodogram peaks at periods longer than the stellar rotation period. As such, additional radial velocity monitoring over future seasons or novel stellar-activity mitigation approaches will be necessary to confirm or refute the candidate second planet signal at $\sim$71 days. Lastly, with no evidence for transits elsewhere in the light-curve, we can attribute the significant LLP peaks interior to planet b as a result of stellar-activity and/or a nightly observing cadence.

\begin{table*}
\caption{Gaussian and min/max priors for quasi-periodic hyper-parameters for TOI-257 used in \texttt{RadVel}. \textbf{Notes.}--(a) These interpretations are further subject to the specific combination of values for the hyper-parameters, notably for cases with significantly different length and timescale factors. See \citet{angus2018} for further discussion. (b) Also used for the initial guess.}
\begin{center}
\begin{minipage}{\linewidth}
\begin{tabularx}{\textwidth}{ccccccc}
\hline\hline
Parameter & Unit; Physical Interpretation$^{a}$ & $\mu^{b}$ & $\sigma$ & Min & Max & Citation \\
\hline
$\eta_{1}$ & \mos, RV amplitude & 10 & None  & 0 & 100  & stddev. of RVs, over-estimate \\
$\eta_{2}$ & days, star-spot decay time-scale & 10  & 5  & 0 & 100 & Estimated from \citet{giles2017}, Fig. 5 \\
$\eta_{3}$ & days, quasi-period & 4.036 ($\times$ 2)  & 0.134 ($\times$ 2) & 0 & 100 & \textit{TESS} light curve; Section \ref{star_rot}, this paper \\
$\eta_{4}$ & none, period length scale & 0.3525 & 0.044 & 0 & 100 & \cite{dai2017, haywood2018} \\
\hline
\end{tabularx}
\end{minipage}
\end{center}
\label{hyperparams}
\end{table*}

\begin{table}
\caption{The relative Bayesian information criteria (BIC) and small sample Akaike information criteria (AICc) \citep{akaike1974} for the various models tested in \texttt{RadVel}. The GP model is marginally favored according to the AICc, but not the BIC. The candidate planet c is highly favored, however additional measurements will be necessary to confirm (or deny) its existence without a transiting event.}
\begin{center}
\begin{minipage}{\linewidth}
\begin{tabularx}{\textwidth}{cccc}
\hline\hline
Model & Number of free parameters & $\Delta$ AICc & $\Delta$ BIC \\
\hline
$b$, $c$, $GP$, $\sigma$ & 18 & 0 & 8.80 \\
$b$, $c$, $\sigma$ & 14 & 2.73 & 0 \\
$b$, $GP$, $\sigma$ & 15 & 21.76 & 21.95  \\
$c$, $GP$, $\sigma$ & 17 & 47.31 & 53.27 \\
$b$, $\sigma$ & 11 & 40.44 & 28.78 \\
$c$, $\sigma$ & 13 & 64.89 & 59.21 \\
$\sigma$, $GP$ & 14 & 54.27 & 51.54 \\
$\sigma$ & 10 & 81.10 & 66.41 \\
\hline
\end{tabularx}
\end{minipage}
\end{center}
\label{tab:info_criteria}
\end{table}

\section{Discussion}
\label{discussion}
Here we have presented the discovery of TOI-257b, the first {\sc {\textsc{Minerva}}}-Australis led confirmation of a \textit{TESS} transiting planet candidate. TOI-257b is a warm sub-Saturn planet with a radius $\sim24\%$ smaller than Saturn ($R_P=7.16\pm0.15$\,\re) and a mass $\sim54\%$ less than Saturn ($M_P=43.9\pm7.3$\,\me) on an orbit of $P=18.38818^{+0.00085}_{-0.00084}$\,$\rm{days}$. The measured mass and radius give a mean density of $0.65^{+0.12}_{-0.11}$\,\densitycgs, consistent with the density of Saturn (0.687\,\densitycgs) and less dense than Jupiter (1.326\,\densitycgs). Therefore, based on the mass, radius, and bulk density of this planet, it lies within the regime of planets classified as `Neptunian worlds' by \citet{chen2017}. Further analysis of the radial velocity data also reveals hints for a second sub-Saturn mass planet ($M_P=70\pm14$\,\me) in the system with an orbit of $\sim71$\,days. However, additional high-precision radial velocity data is required to confirm the planet c candidate.

To understand the planet formation process, we must determine the bulk compositions of warm sub-Saturns such as TOI-257b, a class of planet which is absent from the Solar System. Such objects provide important data for astronomers studying planetary interiors because their masses are sufficiently small that their cores are not degenerate. That is, their mass and radius are dependent on each other such that the core and envelope mass fraction provides single family of solutions \citep[e.g.][]{weiss14, petigura16, pepper17, petigura17}. For planets near the mass of Jupiter, cores are degenerate, and planetary radii are essentially independent of mass. Warm sub-Saturns represent an observational sweet spot where mass and radius are comparatively easy to measure, and when used to interpret the observations, standard models deliver a well-defined family of solutions for the planet's core/envelope mass ratio. This is particularly true for sub-Saturns with incident flux less than the $\sim0.2\times$\,\fluxcgs~ limit where stellar irradiation can inflate planetary radii \citep{demory2011}. The incident flux for TOI-257b is $\sim0.25\times$\,\fluxcgs~ and is very near this limit. Thus the effects of stellar irradiation on the radius of TOI-257b are likely negligible, allowing its internal structure to be modeled and highlights the significant value of discovering other similar planets with low incident flux.

\begin{figure}
  \includegraphics[width=8.5cm]{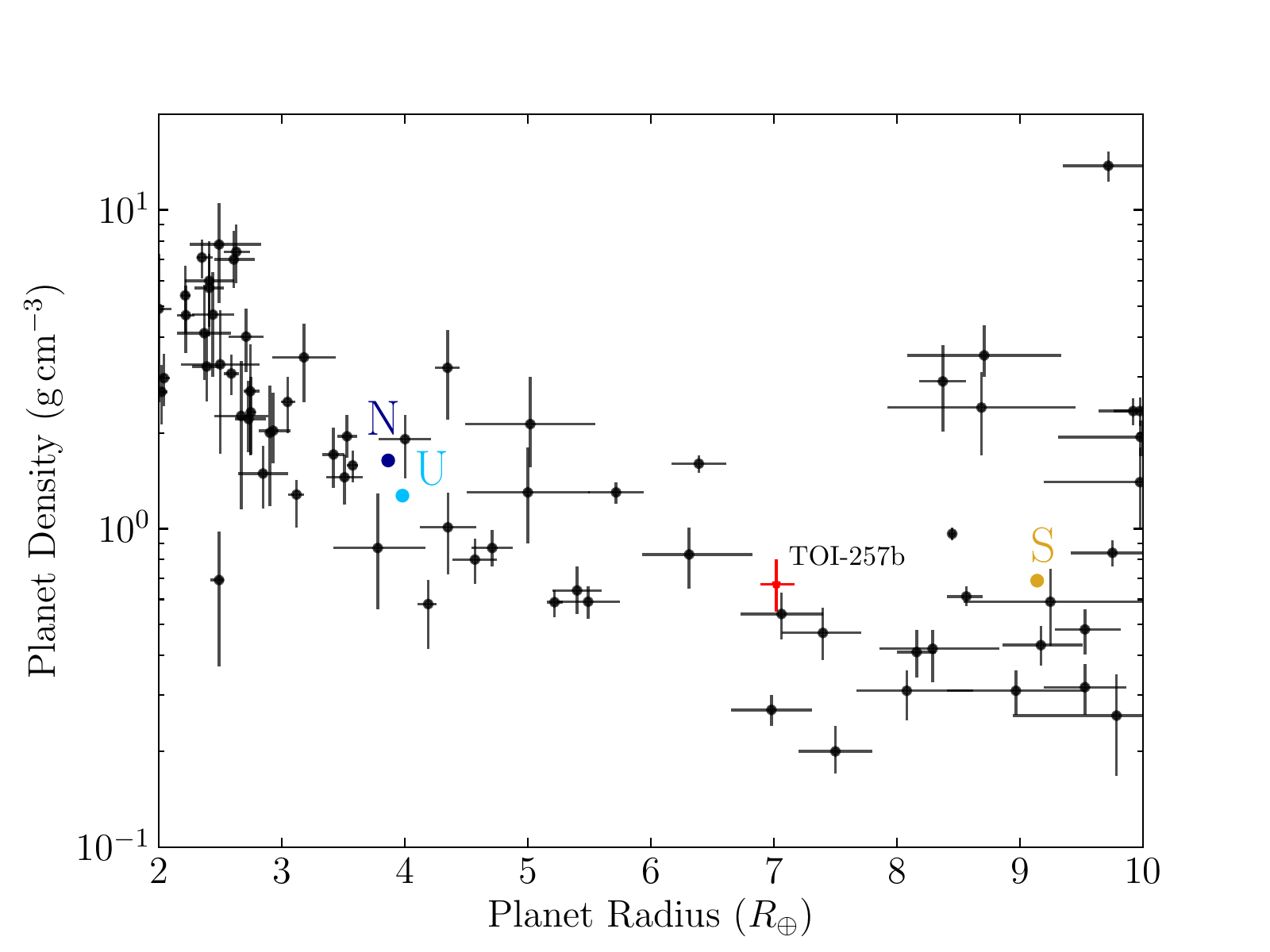}
  \caption{Planet radii versus density for Neptunian planets with $R_{P}=2-10$\,$R_{\oplus}$ and that have a density measured to better than 50\%. TOI-257b studied in this paper is labeled and plotted in red. The Solar System planets Saturn, Uranus, and Neptune are plotted as the gold colored letter S, light blue colored letter U, and dark blue colored letter N, respectively. Planets are taken from the NASA Exoplanet Archive (\url{https://exoplanetarchive.ipac.caltech.edu/}).}
  \label{radius_density_fig}
\end{figure}

\begin{figure}
  \includegraphics[width=8.5cm]{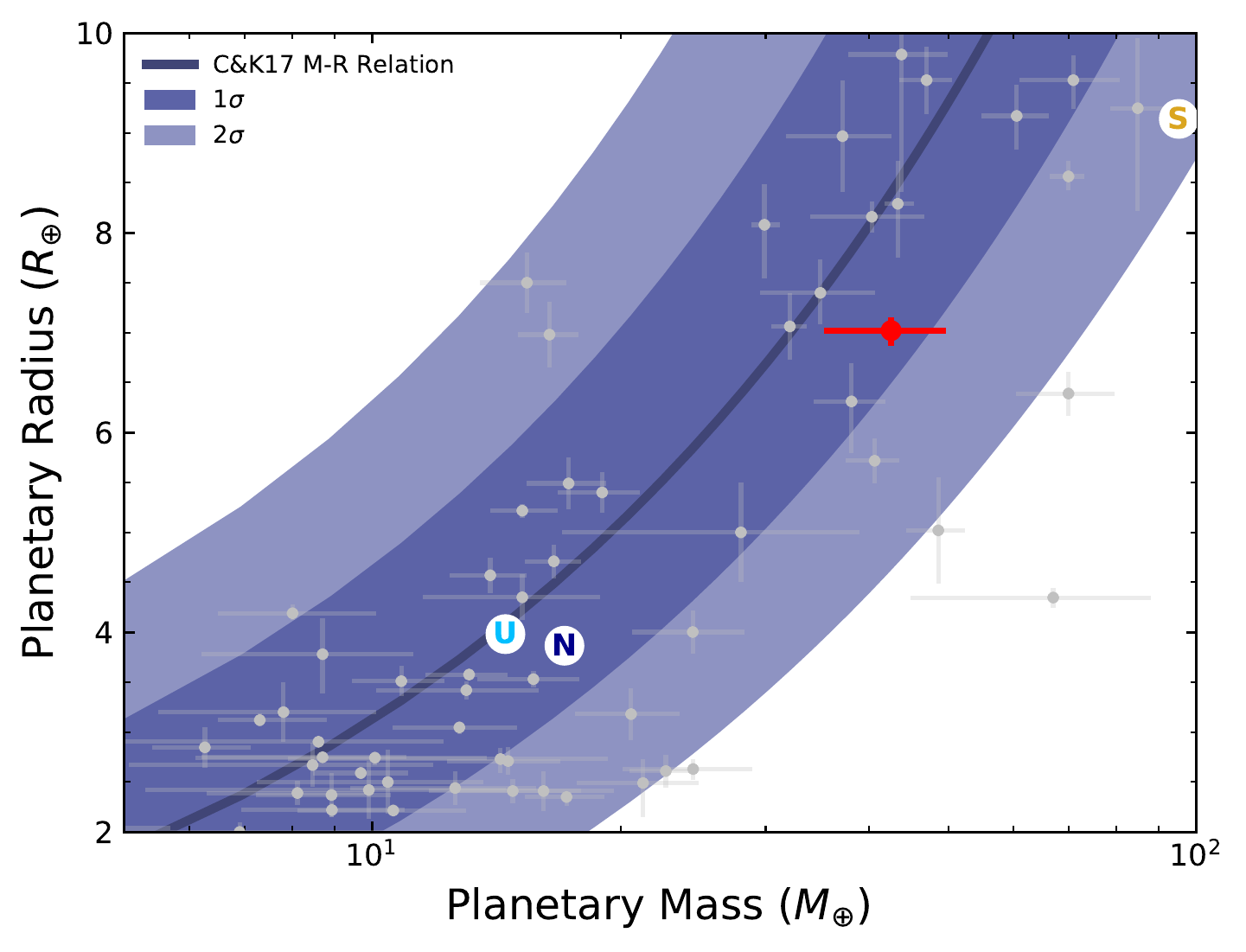}
  \caption{Planet mass versus radius for Neptunian planets with $M_{P}=5-100$\,$M_{\oplus}$ and $R_{P}=2-10$\,$R_{\oplus}$ measured to better than 50\%. The black line shows the \citealt{chen2017} probabilistic mass-radius relation for Neptunian worlds and the surrounding dark and light regions are the associated 68\% and 95\% confidence intervals, respectively. Similar to Figure~\ref{radius_density_fig}, TOI-257b is plotted in red and the Solar System planets Saturn, Uranus, and Neptune are plotted as the gold colored letter S, light blue colored letter U, and dark blue colored letter N, respectively. Planets are taken from the NASA Exoplanet Archive (\url{https://exoplanetarchive.ipac.caltech.edu/}).}
  \label{mass_radius_fig}
\end{figure}

Figure~\ref{radius_density_fig} shows the radius-density diagram for Neptunian worlds (similarly defined after \citealt{chen2017} as those with radii from $\sim2-10$\,\Rearth). This classification also includes mini-Neptunes, sub-Saturns, and Saturns, planets that are dominated by large atmospheres of hydrogen and helium gas and are not significantly effected by gravitational self-compression. We show those planets for which the density has been measured to a precision of better than 50\%. TOI-257b has a mean density that is comparable to other exoplanets around the same size. Figure~\ref{radius_density_fig} also shows the apparent trend of decreasing bulk density as a function of planet radius, indicative of the increasingly large volatile gas envelope up to around the radius of Saturn. Figure~\ref{mass_radius_fig} shows the mass-radius diagram for planets with masses between $5-100$\,\Mearth\ and radii between $2-10$\,\Rearth\ and for which they have been measured to a precision of better than 50\% with the \citealt{chen2017} probabilistic mass-radius relation for Neptunian worlds over-plotted. As evident in Figure~\ref{mass_radius_fig}, TOI-257b lies within the $1\sigma$\, uncertainty region of the mass-radius relationship as predicted by \citealt{chen2017}.

%In Figure~\ref{period_eccentricity_fig} we plot the orbital period versus eccentricity for well-characterized transiting exoplanets with a measured mass from radial velocity measurements. The size of the symbols scales with $\log_{10}$ of the planet mass. In the single planet solution for the system, TOI-257b is on an eccentric orbit of $e=0.242^{+0.040}_{-0.065}$ and lies near the upper range of eccentricity values for other `warm' Neptune and Jovian planets that have orbital periods of $P\geq10$\,days in this sample. Figure~\ref{period_eccentricity_fig} also shows that planets on short period orbits of $P\leq10$\,days tend to have nearly circular orbits, likely due to affects of tidal interactions with the host star circularizing the orbits that were once more eccentric \citep{fabrycky2007}. For planets orbiting beyond $P\sim10$\,days, tidal effects with the host star are expected to be too weak to fully circularize the orbits and a more broad distribution of orbital eccentricity is observed.

The moderately low bulk density ($0.65$\,\densitycgs) and relatively high equilibrium temperature (1027\,K) for this planet as well as it orbiting a bright ($J_{mag}=6.504\pm0.020$ and $K_{mag}=6.256\pm0.020$) host star make it a potentially enticing target for follow-up atmospheric characterisation from the upcoming James Webb Space Telescope (JWST). Using the transmission spectroscopy metric (TSM) of \citet{2018PASP..130k4401K}, we find that this planet has a TSM of $\sim142$. TSMs greater than 90 for Jovian and sub-Jovian planets are considered suitable for JWST transmission spectroscopy observations, making TOI-257\,b an excellent target. However, this planet is not very suitable for emission spectroscopy given that the planet is on a relatively long period orbit and is cool compared to other planets with thermal emission measurements. We estimate that the expected secondary eclipse at 5\,$\mu$m has a depth of $\sim40$\,ppm (assuming blackbody emission), which would be challenging to measure, making this target less suitable for emission spectroscopy.

Measurements of the spin-orbit alignment for transiting warm Neptunian and Jovian worlds via the Rossiter-McLaughlin effect can provide powerful insights into the origins and migration histories of these planets \citep[e.g.,][]{queloz2000,chatterjee2008,winn2010,naoz2011,addison2018,2018AJ....155...70W}. Both classes of planets are strongly believed to have been formed beyond their hosts' protostellar ice line \citep[for a dissenting view of the formation of close-in gas giant planets in-situ via the core-accretion process, see e.g.,][]{2016ApJ...829..114B,2019A&A...629L...1H} and then experienced inward migration through one of two types of migration channels, quiescent migration through the disk \citep{lin1996} or chaotically dynamical high-eccentricity migration \citep{fabrycky2007,ford2008c,naoz2011}. The latter migration mechanism is thought to be responsible for producing many of the known hot Jupiters due to the large observed range in their spin-orbit angles \citep[e.g., see,][]{albrecht2012,addison2013,addison2018}. However, it is unknown if this is the case for the warm sub-Saturn and Neptunian worlds like TOI-257b with orbits greater than 10\,days. The limited sample of spin-orbit angles measured for these planet populations (only seven so far according to the TEPCat catalog\footnote{\url{https://www.astro.keele.ac.uk/jkt/tepcat/}}, see \citealt{2011MNRAS.417.2166S}) makes it difficult to draw any firm conclusions and more measurements are urgently needed. This planet presents a suitable candidate for studying the spin-orbit via the Rossiter-McLaughlin effect. We predict that the radial velocity semi-amplitude of the Rossiter-McLaughlin effect for TOI-257 to be $\sim8$\,\mos~ based on the stellar and planetary parameters we obtained for this system. The predicted signal, while small, should be detectable on very high-precision ($\sim1$\,\mos) radial velocity facilities in the south such as on HARPS \citep{2004SPIE.5492..148R}, PFS \citep{2006SPIE.6269E..31C}, and ESPRESSO \citep{2010SPIE.7735E..0FP}. We predict, given the stellar rotational velocity (as determined from the rotational period and stellar radius) is consistent with the measured $v\sin i$ from spectroscopy (i.e., suggesting that the stellar obliquity is near $90\deg$), that the projected spin-orbit angle $\lambda$ when measured (whether aligned or misaligned) should be close to the true spin-orbit angle $\psi$.

%\begin{figure}
%  \includegraphics[width=8.5cm]{period_vs_eccentricity.pdf}
%  \caption{Orbital period versus eccentricity for well-characterized transiting exoplanets and confirmed by radial velocities. TOI-257b is labeled and plotted in red. The size of the plotted circles scales with $\log_{10}$ of the planet mass. Circles in black are those with a measured uncertainty for the eccentricity while those in orange have no reported uncertainty on the eccentricity value.}
%  \label{period_eccentricity_fig}
%\end{figure}

\section{Conclusions}
\label{conclusions}
We report the discovery of TOI-257b, a $R_P=0.639\pm0.013$\,$\rm{R_J}$ ($R_{P}=7.16$\,\re) and $M_P=0.138\pm0.023$\,$\rm{M_J}$ ($M_{P}=43.9$\,\me) transiting planet found by \textit{TESS} and confirmed using radial velocity data from {\textsc{Minerva}}-Australis, FEROS, and HARPS as well as direct imaging from SOAR and Zorro. We also find hints for an additional non-transiting long-period ($\sim71$\,day) sub-Saturn mass planet candidate orbiting TOI-257 from analysis of the radial velocity data. TOI-257b belongs to a population of exoplanets between the sizes of Neptune and Saturn that appears to be rare. Furthermore, TOI-257b transits a very bright star ($V = 7.612$\,mag) on a relatively long-period orbit of 18.423\,days making it a great candidate for future follow-up observations to measure its spin-orbit alignment and transmission spectrum. Warm sub-Saturns such as TOI-257\,b are important population of planets to study for understanding the formation, internal structures and compositions, and evolution and migration of giant planets. Future observational work of this planetary system will help to elucidate our understanding of these rare sub-Saturn planets that are absent in the Solar System.

%to probe its atmospheric composition through transmission and emissions spectroscopy and

%%%%%%%%%%%%%%%%%%%%%%%%%%%%%%%%%%%%%%%%%%%%%%%%%%

\section*{Data Availability Statements}
The radial velocity data underlying this article are available in the article. The \textit{TESS} photometric data is available at NASA's Mikulski Archive for Space Telescopes (\url{https://archive.stsci.edu/}). The raw direct imaging data and raw spectra will be shared on reasonable request to the corresponding author.

\section*{Acknowledgements}

We thank the anonymous referee for their careful review of the manuscript that has led to significant improvements.

{\textsc{Minerva}}-Australis is supported by Australian Research Council LIEF Grant LE160100001, Discovery Grant DP180100972, Mount Cuba Astronomical Foundation, and institutional partners University of Southern Queensland, UNSW Australia, MIT, Nanjing University, George Mason University, University of Louisville, University of California Riverside, University of Florida, and The University of Texas at Austin.

We respectfully acknowledge the traditional custodians of all lands throughout Australia, and recognise their continued cultural and spiritual connection to the land, waterways, cosmos, and community. We pay our deepest respects to all Elders, ancestors and descendants of the Giabal, Jarowair, and Kambuwal nations, upon whose lands the {\textsc{Minerva}}-Australis facility at Mt Kent is situated.

B.P.B. acknowledges support from the National Science Foundation grant AST-1909209.

This research has made use of the NASA Exoplanet Archive, which is operated by the California Institute of Technology, under contract with the National Aeronautics and Space Administration under the Exoplanet Exploration Program. Funding for the \textit{TESS} mission is provided by NASA's Science Mission directorate. We acknowledge the use of public \textit{TESS} Alert data from pipelines at the \textit{TESS} Science Office and at the \textit{TESS} Science Processing Operations Center. The results reported herein benefited from collaborations and/or information exchange within NASA's Nexus for Exoplanet System Science (NExSS) research coordination network sponsored by NASA's Science Mission Directorate. Based on observations obtained at the Gemini Observatory, which is operated by the Association of Universities for Research in Astronomy, Inc., under a cooperative agreement with the NSF on behalf of the Gemini partnership: the National Science Foundation (United States), National Research Council (Canada), CONICYT (Chile), Ministerio de Ciencia, Tecnolog\'{i}a e Innovaci\'{o}n Productiva (Argentina), Minist\'{e}rio da Ci\^{e}ncia, Tecnologia e Inova\c{c}\~{a}o (Brazil), and Korea Astronomy and Space Science Institute (Republic of Korea). Some of the Observations in the paper made use of the High-Resolution Imaging instrument Zorro at Gemini-South. Zorro was funded by the NASA Exoplanet Exploration Program and built at the NASA Ames Research Center by Steve B. Howell, Nic Scott, Elliott P. Horch, and Emmett Quigley. This research has made use of NASA's Astrophysics Data System.

D.H.\ acknowledges support by the National Aeronautics and Space Administration through the \textit{TESS} Guest Investigator Program (80NSSC18K1585) and by the National Science Foundation (AST-1717000). 
A.C.\ acknowledges support by the National Science Foundation under the Graduate Research Fellowship Program.
W.J.C., W.H.B., M.B.N. and A.M.\ acknowledge support from the Science and Technology Facilities Council and UK Space Agency. Funding for the Stellar Astrophysics Centre is provided by The Danish National Research Foundation (Grant DNRF106).
R.B.\ acknowledges support from FONDECYT Post-doctoral Fellowship Project 3180246, and from the Millennium Institute of Astrophysics (MAS).
H.Z.\ Hui Zhang is supported by the Natural Science Foundation of China ( NSFC grants 11673011, 11933001).
A.J.\ acknowledges support from FONDECYT project 1171208 and by the Ministry for the Economy, Development, and Tourism's Programa Iniciativa Cient\'{i}fica Milenio through grant IC\,120009, awarded to the Millennium Institute of Astrophysics (MAS). 
A.M.S.\ is partially supported by grants ESP2017-82674-R (Spanish Government) and 2017-SGR-1131 (Generalitat de Catalunya).
A.M.\ acknowledges support from the ERC Consolidator Grant funding scheme (project ASTEROCHRONOMETRY, G.A. n. 772293).
R.A.G.\ acknowledge the support of the PLATO grant from the CNES. 
%S.M.\ acknowledges support from the European Research Council through the SPIRE grant 647383.
S.M.\ acknowledges support from the Spanish Ministry with the Ramon y Cajal fellowship number RYC-2015-17697. 
T.L.C.\ acknowledges support from the European Union's Horizon 2020 research and innovation programme under the Marie Sk\l{}odowska-Curie grant agreement No.~792848 (PULSATION). This work was supported by FCT/MCTES through national funds (UID/FIS/04434/2019).
E.C.\ is funded by the European Union’s Horizon 2020 research and innovation program under the Marie Sklodowska-Curie grant agreement No. 664931.
V.S.A.\ acknowledges support from the Independent Research Fund Denmark (Research grant 7027-00096B).
S.B.\ acknowledges NASA grant NNX16AI09G and NSF grant AST-1514676.
I.J.M.C.\ acknowledges support from the NSF through grant AST-1824644, and from NASA through Caltech/JPL grant RSA-1610091.
T.D.\ acknowledges support from MIT’s Kavli Institute as a Kavli postdoctoral fellow.
D.B.\ acknowledges support from NASA through the TESS GI program (80NSSC19K0385).
C.K.\ acknowledges support by Erciyes University Scientific Research Projects Coordination Unit under grant number MAP-2020-9749.
E.L. and M.C.\ acknowledge support by the National Science Foundation under grant 1559487.
V.S.A.\ acknowledges support from the Independent Research Fund Denmark (Research grant 7027-00096B) and the Carlsberg foundation (grant agreement CF19-0649).

\smallskip

\textit{\Large{Software:}} Astropy \citep{astropy}, Matplotlib \citep{matplotlib}, DIAMONDS \citep{corsaro14}, isoclassify \citep{huber17}, \texttt{EXOFASTv2} \citep{2013PASP..125...83E,2017ascl.soft10003E,2019arXiv190709480E}

%%%%%%%%%%%%%%%%%%%%%%%%%%%%%%%%%%%%%%%%%%%%%%%%%%



\bibliographystyle{mnras}
\bibliography{references} % if your bibtex file is called example.bib

%\begin{thebibliography}{99}
%\end{thebibliography}

%%%%%%%%%%%%%%%%%%%%%%%%%%%%%%%%%%%%%%%%%%%%%%%%%%

%%%%%%%%%%%%%%%%% APPENDICES %%%%%%%%%%%%%%%%%%%%%

\clearpage
\newpage

\appendix

\section{\texttt{RadVel} Posterior Distribution Plots}

\begin{figure}
  \centering
  \includegraphics[width=\textwidth]{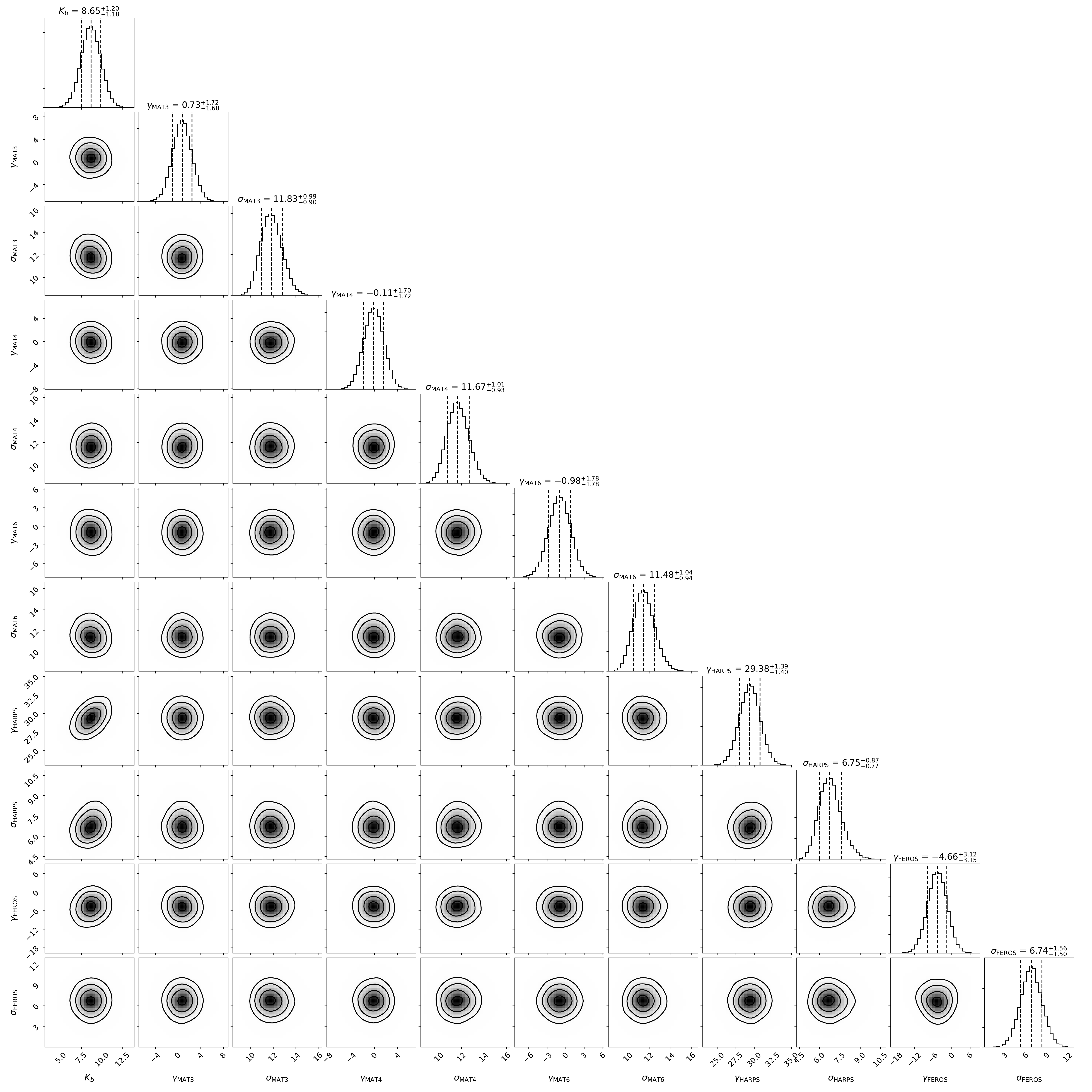}
  \caption{Posterior distributions from \texttt{RadVel} for all parameters for a 1-planet circular model.}
  \label{fig:radvel_4}
\end{figure}

\clearpage
\newpage

\begin{figure}
  \includegraphics[width=\textwidth]{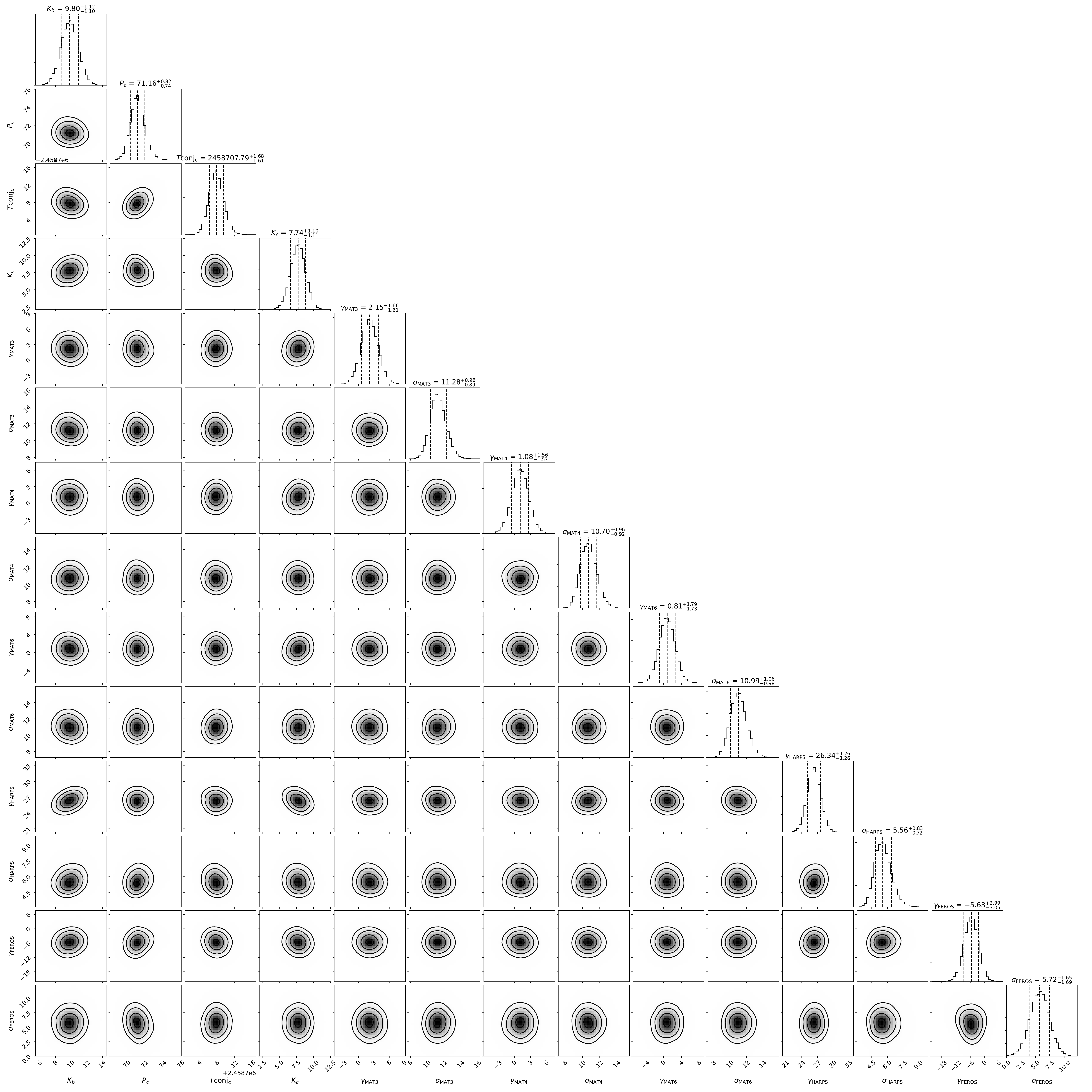}
  \caption{Same as Fig. \ref{fig:radvel_4}, but for a 2-planet circular model.}
  \label{fig:radvel_5}
\end{figure}

\clearpage
\newpage

\begin{figure}
  \includegraphics[width=\textwidth]{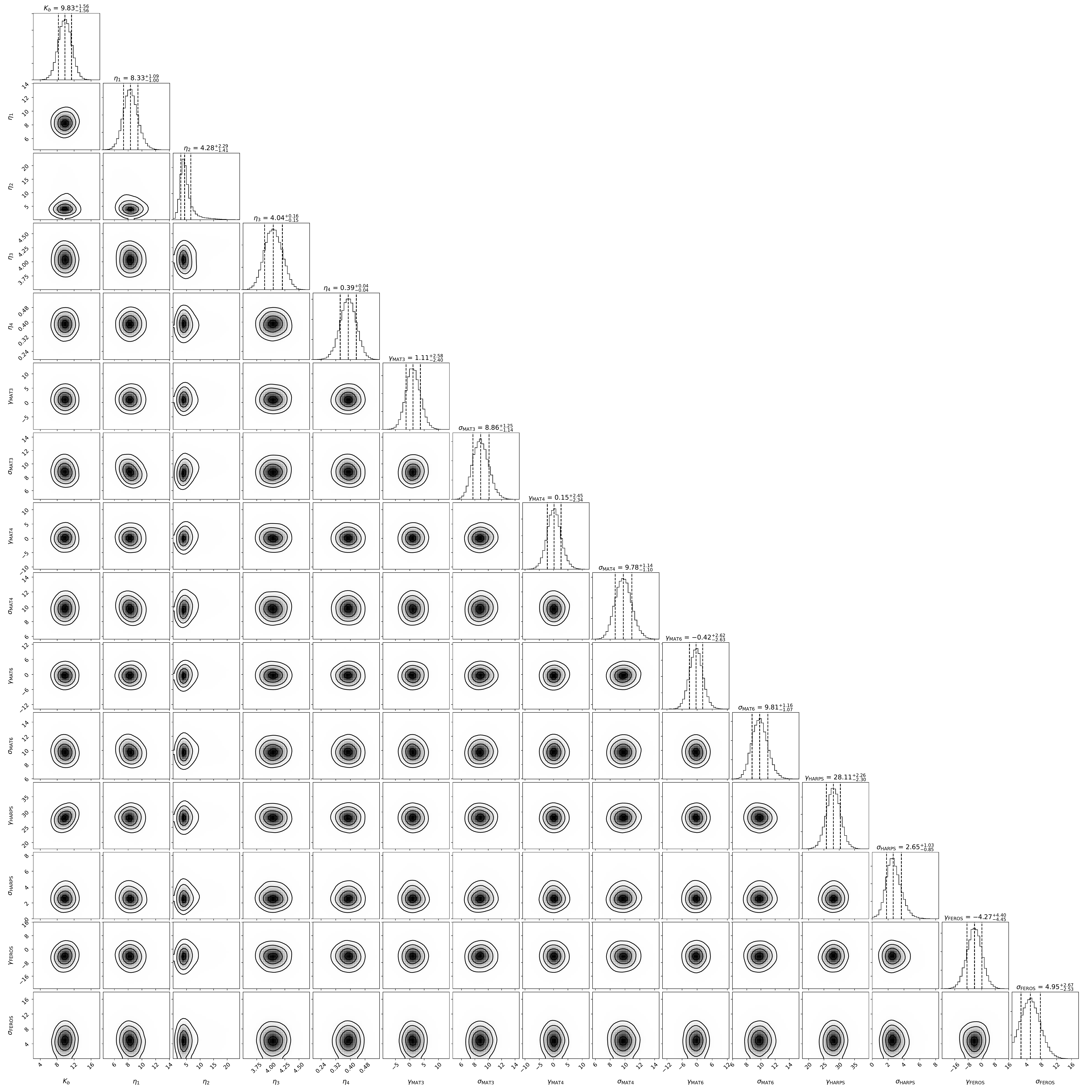}
  \caption{Posterior distributions for all parameters for a 1-planet circular model in \texttt{RadVel} with a quasi-periodic Gaussian Process to model stellar-activity. Priors for hyper-parameters are provided in Table~\ref{hyperparams}.}
  \label{fig:radvel_6}
\end{figure}

\clearpage
\newpage

\begin{figure}
  \includegraphics[width=\textwidth]{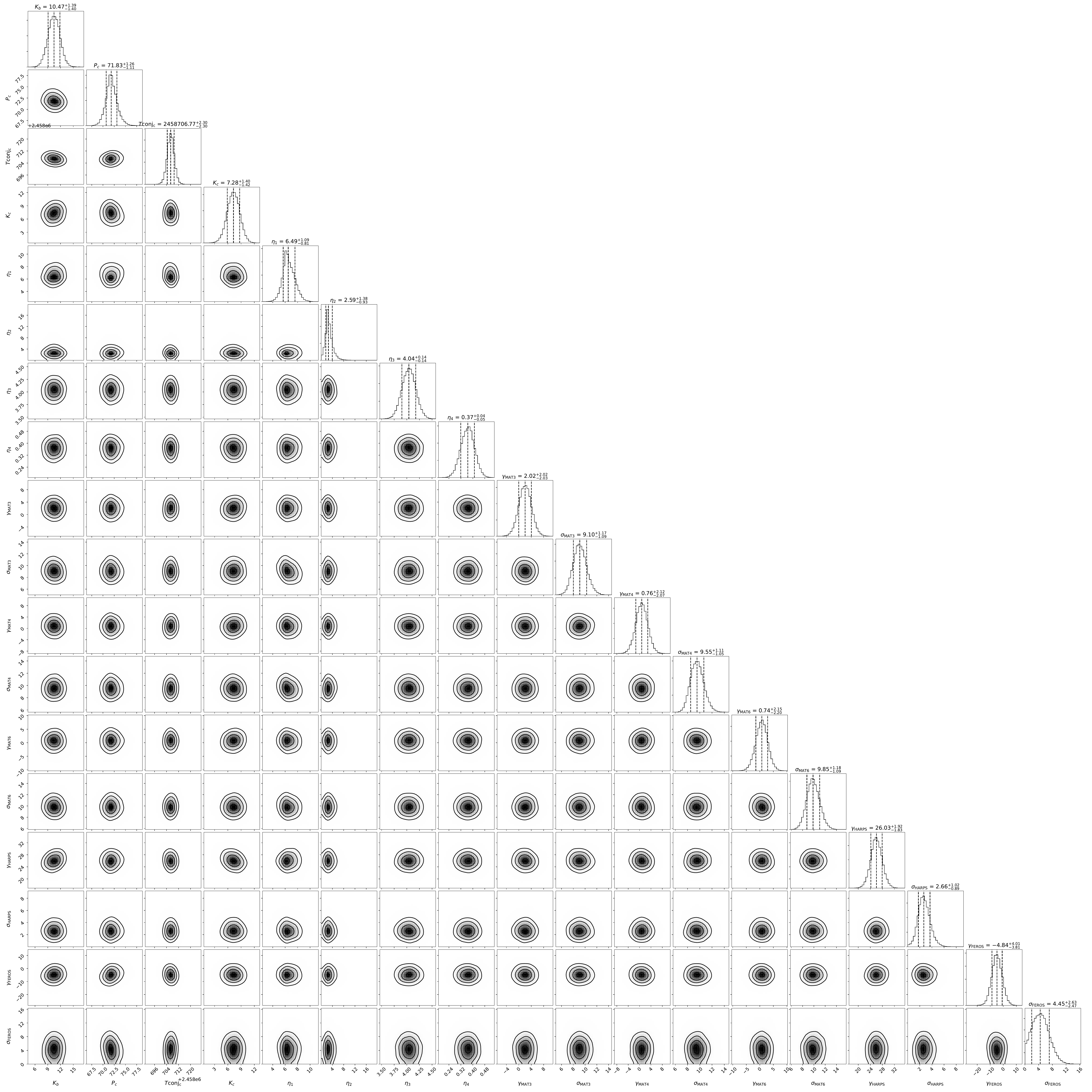}
  \caption{Same as Fig. \ref{fig:radvel_6}, but for a 2-planet circular model.}
  \label{fig:radvel_7}
\end{figure}

\clearpage
\newpage

\section{Author Affiliations}
\label{sec:affiliations}
$^{1}$University of Southern Queensland, Centre for Astrophysics, West Street, Toowoomba, QLD 4350 Australia\\
$^{2}$Sub-department of Astrophysics, Department of Physics, University of Oxford, Denys Wilkinson Building, Keble Road, Oxford, OX1 3RH, UK\\
$^{3}$Department of Physics \& Astronomy, George Mason University, 4400 University Drive MS 3F3, Fairfax, VA 22030, USA\\
$^{4}$Department of Earth and Planetary Sciences, University of California, Riverside, CA 92521, USA\\
$^{5}$Institute for Astronomy, University of Hawai`i, 2680 Woodlawn Drive, Honolulu, HI 96822, USA\\
$^{6}$Department of Astronomy, University of Wisconsin-Madison, Madison, WI, USA 53706\\
$^{7}$School of Physics and Astronomy, University of Birmingham, Birmingham B15 2TT, UK\\
$^{8}$Stellar Astrophysics Centre (SAC), Department of Physics and Astronomy, Aarhus University, Ny Munkegade 120, DK-8000 Aarhus C, Denmark\\
$^{9}$NSF Graduate Research Fellow\\
$^{10}$Center for Astrophysics ${\rm \mid}$ Harvard {\rm \&} Smithsonian, 60 Garden Street, Cambridge, MA 02138, USA\\
$^{11}$Dunlap Institute for Astronomy and Astrophysics, University of Toronto, 50 St. George Street, Toronto, Ontario M5S 3H4, Canada\\
$^{12}$Facultad de Ingeniería y Ciencias, Universidad Adolfo Ib\'a\~nez, Av.\ Diagonal las Torres 2640, Pe\~nalol\'en, Santiago, Chile\\
$^{13}$Millennium Institute for Astrophysics, Faculty of Physics, Campus San Joaquin UC, Av. Vicuna Mackenna, 4860, Mascul, Santiago, Chile\\
$^{14}$Space Telescope Science Institute, 3700 San Martin Drive, Baltimore, MD 21218, USA\\
$^{15}$Astronomy Department, Indiana University Bloomington, 727 East 3rd Street, Swain West 318, IN 4740, USA\\
$^{16}$Facultad de Ingeniera y Ciencias, Universidad Adolfo Ib\'a\~nez, Av. Diagonal las Torres 2640, Pe\~nalol\'en, Santiago, Chile\\
$^{17}$Department of Physics and Astronomy, University of Louisville, Louisville, KY 40292, USA\\
$^{18}$Department of Physical Sciences, Kutztown University, Kutztown, PA 19530, USA\\
$^{19}$Vanderbilt University, Department of Physics \& Astronomy, 6301 Stevenson Center Ln., Nashville, TN 37235, USA\\
$^{20}$Fisk University, Department of Physics, 1000 18th Ave. N., Nashville, TN 37208, USA\\
$^{21}$Sydney Institute for Astronomy (SIfA), School of Physics, University of Sydney, 2006, Australia\\
$^{22}$Department of Astronomy, The University of Texas at Austin, Austin, TX 78712, USA\\
$^{23}$Dept. of Physics and Astronomy, University of Kansas, 1251 Wescoe Hall Dr.,Lawrence, KS 66045, USA\\
$^{24}$Department of Physics, Massachusetts Institute of Technology, Cambridge, MA, USA\\
$^{25}$Department of Astronomy, Yale University, New Haven, CT 06511, USA\\
$^{26}$U.S. Naval Observatory, 3450 Massachusetts Avenue NW, Washington, D.C. 20392, USA\\
$^{27}$Space Science \& Astrobiology Division, NASA Ames Research Center, Moffett Field, CA 94035, USA\\
$^{28}$Department of Physics and Astronomy, The University of North Carolina at Chapel Hill, Chapel Hill, NC 27599-3255, USA\\
$^{29}$Instituto de Astrof\'isica, Pontificia Universidad Cat\'olica de Chile, Av. Vicu\~na Mackenna 4860, Macul, Santiago, Chile\\
$^{30}$Department of Physics, Westminster College, 319 South Market Street, New Wilmington, PA 16172, USA\\
$^{31}$MIT Kavli Institute for Astrophysics and Space Research, Massachusetts Institute of Technology, Cambridge, MA 02139, USA\\
$^{32}$Exoplanetary Science at UNSW, School of Physics, UNSW Sydney, NSW 2052, Australia\\
$^{33}$School of Astronomy and Space Science, Key Laboratory of Modern Astronomy and Astrophysics in Ministry of Education, Nanjing University, Nanjing 210046, Jiangsu, China\\
$^{34}$Max-Planck-Institut f\"ur Astronomie, K\"onigstuhl 17, Heidelberg 69117, Germany\\
$^{35}$Earth and Planetary Sciences, Massachusetts Institute of Technology, 77 Massachusetts Avenue, Cambridge, MA 02139, USA\\
$^{36}$Department of Astrophysical Sciences, Princeton University, 4 Ivy Lane, Princeton, NJ 08544, USA\\
$^{37}$NASA Ames Research Center, Moffett Field, CA 94035, USA\\
$^{38}$School of Physical Sciences, The Open University, Milton Keynes MK7 6AA, UK\\
$^{39}$Department of Physics, Lehigh University, 16 Memorial Drive East, Bethlehem, PA 18015, USA\\
$^{40}$Exoplanets and Stellar Astrophysics Laboratory, Mail Code 667, NASA Goddard Space Flight Center, Greenbelt, MD 20771, USA\\
$^{41}$Perth Exoplanet Survey Telescope, Perth, Western Australia\\
$^{42}$Center for Space Science, NYUAD Institute, New York University Abu Dhabi, PO Box 129188, Abu Dhabi, United Arab Emirates\\
$^{43}$Department of Chemistry \& Physics, Florida Gulf Coast University, 10501 FGCU Blvd. S., Fort Myers, FL 33965 USA\\
$^{44}$Instituto de Astrof\'isica e Ci\^encias do Espa\c{c}o, Universidade do Porto, CAUP, Rua das Estrelas, 4150-762 Porto, Portugal\\
$^{45}$Departamento de F\'{\i}sica e Astronomia, Faculdade de Ci\^{e}ncias da Universidade do Porto, Rua do Campo Alegre, s/n, PT4169-007 Porto, Portugal\\
$^{46}$INAF - Osservatorio Astrofisico di Catania, via S. Sofia 78, 95123, Catania, Italy\\
$^{47}$Instituto de Astrof\'isica de Canarias (IAC), 38205 La Laguna, Tenerife, Spain\\
$^{48}$Universidad de La Laguna (ULL), Departamento de Astrof\'isica, E-38206 La Laguna, Tenerife, Spain\\
$^{49}$Universidade Federal do Rio Grande do Norte (UFRN), Departamento de F\'isica, 59078-970, Natal, RN, Brazil\\
$^{50}$IRFU, CEA, Universit\'e Paris-Saclay, F-91191 Gif-sur-Yvette, France\\
$^{51}$AIM, CEA, CNRS, Universit\'e Paris-Saclay, Universit\'e Paris Diderot, Sorbonne Paris Cit\'e, F-91191 Gif-sur-Yvette, France\\
$^{52}$Center for Exoplanets and Habitable Worlds, Department of Astronomy \& Astrophysics, 525 Davey Laboratory, The Pennsylvania State University, University Park, PA 16802, USA\\
$^{53}$Max-Planck-Institut f{\"u}r Sonnensystemforschung, Justus-von-Liebig-Weg 3, 37077 G{\"o}ttingen, Germany\\
$^{54}$Institute of Astrophysics, University of Vienna, 1180 Vienna, Austria\\
$^{55}$Department of Physics and Astronomy, Iowa State University, Ames, IA 50011 USA\\
$^{56}$Department of Astronomy \& Space Sciences, Erciyes University, Kayseri, Turkey\\
$^{57}$LESIA, Observatoire de Paris, Universit\'e PSL, CNRS, Sorbonne Universit\'e, Universit\'e de Paris, 92195 Meudon, France\\
$^{58}$Institute of Space Sciences (ICE, CSIC) Campus UAB, Carrer de Can Magrans, s/n, E-08193, Barcelona, Spain\\
$^{59}$Institut d’Estudis Espacials de Catalunya (IEEC), C/Gran Capita, 2-4, E-08034, Barcelona, Spain\\
$^{60}$Zentrum fur Astronomie (ZAH/LSW), University of Heidelberg, Albert-Ueberle-Str. 2, D-69120 Heidelberg, Germany\\
$^{61}$HITS gGmbH, Schloss-Wolfsbrunnenweg 35, 69118 Heidelberg, Germany\\
$^{62}$MIT Department of Physics, Massachusetts Institute of Technology, Cambridge, MA 02139, USA

%%%%%%%%%%%%%%%%%%%%%%%%%%%%%%%%%%%%%%%%%%%%%%%%%%

% Don't change these lines
\bsp	% typesetting comment
\label{lastpage}
\end{document}